\documentclass[12pt,a4paper]{article}
\usepackage[pdftex]{graphicx}
\usepackage{amsmath,amsfonts,amsthm,
lscape,amssymb,mathrsfs,cases}
\usepackage{cite}

\usepackage[normalem]{ulem}
\usepackage[svgnames]{xcolor}
\usepackage{tikz}
\usepackage{textcomp}

\newcommand{\vt}[1]{\mbox{\boldmath$#1$}}
\newcommand{\sca}[2]{\langle #1, #2 \rangle}

\numberwithin{equation}{section}

\def\beqa{\begin{eqnarray}}
\def\enqa{\end{eqnarray}}
\def\beq{\begin{equation}}
\def\enq{\end{equation}}

\begin{document}
\title{
\vspace{-9mm}
Integrable 
discretization
of the 
vector/matrix 
nonlinear Schr\"odinger 
equation
and 
the associated 
Yang--Baxter map
}
\author{Takayuki \textsc{Tsuchida}
}
\maketitle
\begin{abstract} 
%
The action of a 
B\"acklund--Darboux
transformation 
on a 
spectral 
problem 
associated with 
a known integrable system 
can define 
a 
new 
discrete 
spectral problem.  
In this paper, 
we 
interpret 
a 
slightly 
generalized 
version of 
the 
binary 
B\"acklund--Darboux 
(or
Zakharov--Shabat
dressing) 
transformation 
for the nonlinear Schr\"odinger (NLS) hierarchy 
as a 
discrete spectral problem, 
wherein 
the two intermediate 
potentials appearing in 
the 
Darboux 
matrix 
are considered 
as 
a pair of new 
dependent variables. 
Then, we 
associate the 
discrete spectral problem 
with 
a suitable 
isospectral 
time-evolution 
equation, 
which 
forms the Lax-pair representation
for a 
space-discrete 
NLS 
system. 
This formulation is valid for 
the 
most 
general case 
where 
the  
two 
dependent 
variables 
take 
values in 
(rectangular) 
matrices. 
In contrast to the matrix 
generalization of 
the Ablowitz--Ladik 
lattice, 
our discretization has a rational nonlinearity 
and
admits 
a 
Hermitian conjugation reduction 
between the two dependent variables. 
Thus, 
a new 
proper space-discretization of 
the 
vector/matrix 
NLS 
equation 
is obtained; 
by 
changing 
the time part of the Lax pair, 
we 
also 
obtain 
an integrable space-discretization of 
the 
vector/matrix 
modified KdV (mKdV) 
equation. 
Because 
B\"acklund--Darboux transformations 
are permutable, 
we can 
increase the number of discrete independent variables  
in a multi-dimensionally consistent 
way. 
By solving the consistency condition on the two-dimensional lattice, 
we obtain 
a 
Yang--Baxter map of the NLS type, 
which 
can be considered as a 
fully discrete 
analog of 
the principal chiral model for 
projection matrices. 
\end{abstract}
%
\newpage
\noindent
\tableofcontents

\newpage
\section{Introduction}
The problem of 
integrable discretization~\cite{Suris03} 
has a close 
relationship 
with 
the theory of 
auto-B\"acklund transformations; 
this 
fact has been 
sporadically 
noticed 
for 
partial differential 
or differential-difference
equations since the mid-1970s.
Calogero and Degasperis~\cite{Calo75,Calo76}
and 
Chiu and Ladik~\cite{Chiu77} 
showed 
for some specific examples 
that 
a class of auto-B\"acklund transformations 
can be 
identified 
as 
discrete-time 
flows 
that belong to 
the same 
integrable hierarchy as the 
original 
continuous-time flows. 
Hirota~\cite{Hiro77} 
and 
Orfanidis~\cite{Orfa1,Orfa2} 
showed 
for the sine-Gordon equation in light-cone coordinates 
that 
a 
one-parameter auto-B\"acklund transformation
and the associated 
nonlinear superposition principle 
(Bianchi's permutability theorem) 
lead 
directly 
to 
its 
integrable 
discretizations. 
These results 
have 
culminated in the 
beautiful 
notion of 
Miwa variables (or Miwa shifts)
in Sato theory~\cite{Miwa82,Date1}. 
In rough terms, 
a one-parameter 
auto-B\"acklund transformation, 
called an elementary B\"acklund transformation, 
can be reinterpreted 
as a 
discrete-time 
flow, which   
can generate
an 
infinite number 
of 
continuous-time flows 
through 
the Taylor series expansion 
in the 
B\"acklund 
(or 
time-step) 
parameter; 
the 
commutativity 
of 
two 
elementary B\"acklund transformations 
with 
(generally) 
different values of the 
B\"acklund 
parameter
provides 
a fully discrete 
integrable system 
that 
admits a 
zero-curvature (or Lax-pair~\cite{Lax}) representation 
and 
contains 
information on 
the 
continuous-time 
flows. 
%
Such a unified point of view 
is 
so fascinating that 
%
one is tempted to 
believe 
that 
proper 
discretizations 
of a given 
integrable 
system 
can always be obtained 
from 
its 
elementary 
B\"acklund transformation 
or the associated nonlinear superposition formula. 
However, in fact, 
it is not that simple; 
we 
remark 
the following 
points. 
\begin{itemize} 
\item An integrable system 
appears as a member of 
an infinite hierarchy 
of commuting flows, 
which is actually a bi-infinite hierarchy 
comprising the positive 
and 
negative flows. 
For example, 
the sine-Gordon equation 
is the first negative flow 
of the integrable hierarchy, 
while 
its 
first nontrivial positive flow 
is  
(the potential form of) 
the modified KdV (mKdV) equation~\cite{AKNS73,AKNS74}. 
It is
not evident 
which particular 
flows in 
a bi-infinite 
hierarchy can be 
discretized by considering 
an elementary 
B\"acklund transformation
and how to 
take the continuous limit. 
\item 
The idea to 
interpret 
an elementary B\"acklund transformation 
and 
the associated nonlinear superposition formula 
as defining 
discrete 
integrable 
systems 
is 
useful 
for 
integrable 
hierarchies 
with one scalar 
unknown. 
However, 
it is
not 
the case for 
two
(or more) component 
systems
that admit 
a complex conjugation reduction 
between 
a pair of 
dependent variables, 
such as the nonlinear Schr\"odinger (NLS) system~\cite{AKNS73,AKNS74
}, 
as well as their matrix generalizations 
that admit a Hermitian conjugation reduction 
between a pair of matrix dependent variables, 
such as the matrix NLS system~\cite{ZS74}. 
Indeed, 
an 
elementary 
B\"acklund transformation 
generally does 
not
maintain 
the complex/Hermitian conjugation reduction; 
thus, 
%
for a given 
integrable 
equation 
obtained as the 
result of 
the 
reduction, 
we cannot 
derive 
its 
proper 
discretization 
directly from 
an elementary 
B\"acklund transformation 
for the original nonreduced system 
(cf.~\cite{Kono82,Chud1,Chud2,DJM83}). 
Actually, 
we can 
consider a suitable composition of 
two (or more) 
elementary 
B\"acklund transformations~\cite{Kono79,Kono82,Konop3} 
so that the composite ({\it e.g.}, binary)
B\"acklund 
transformation 
can 
maintain 
the complex/Hermitian conjugation reduction; 
however, 
such a composite 
transformation 
involves either 
a square-root function 
with an indefinite sign~\cite{Chen1,Lamb74} 
or a nonlocal operation such as an 
indefinite integral~\cite{Calo76}, 
so it does not lead to 
discrete integrable systems 
that can be written in local form 
and 
define 
a unique time evolution 
(cf.~(7.18) in~\cite{QNCL} 
and 
(4.17) in~\cite{Linden2}). 
\end{itemize}
Thus, for each integrable system 
that admits the complex/Hermitian conjugation reduction, 
we 
need 
to 
construct its proper 
discretization 
on a case-by-case consideration 
and accumulate more 
knowledge 
on this subject. 

In this paper, 
we 
construct 
a 
new 
proper space-discretization of 
the 
(generally rectangular) 
matrix 
generalization of the 
NLS system, 
namely, the matrix NLS system~\cite{ZS74}, 
which  
is an integrable system admitting 
the 
Hermitian conjugation reduction 
between the two matrix
dependent variables  
and 
contains the vector NLS equation 
known as 
the Manakov model~\cite{Mana74}
as a special case. 
For the scalar NLS system, 
the proper and 
most natural 
space-discretization 
was 
found by Ablowitz and Ladik~\cite{AL1,AL76}  
in the mid-1970s; 
however, 
its straightforward matrix generalization~\cite{GI82}
(also see~\cite{Tsuchi02,DM2010} and references therein)
does not 
admit 
a Hermitian conjugation reduction between 
the two 
matrix 
variables in local form, 
so it 
is not 
a 
proper 
space-discretization of the 
matrix NLS
system. 
To obtain a proper 
space-discrete matrix NLS system, 
we follow 
the 
new 
approach 
introduced 
in our
previous 
paper~\cite{Me2015}. 
That is, 
we 
first 
reinterpret 
(a slightly generalized 
version 
of) 
a binary 
B\"acklund--Darboux transformation for the continuous 
matrix NLS hierarchy 
as a discrete spectral problem~\cite{Chud2}, where
the two intermediate potentials appearing in the binary 
B\"acklund--Darboux transformation 
are 
considered as 
new dependent variables. 
Then, in view of the peculiar structure of the discrete spectral problem, 
we associate 
it with a suitable isospectral time-evolution equation 
to compose a Lax pair 
%
and derive an evolutionary lattice system 
from the compatibility condition 
called the zero-curvature equation. 
This lattice system involves some free parameters 
and, with a suitable choice of the parameters, 
it 
admits 
the Hermitian conjugation reduction 
between the two dependent variables 
and 
provides a proper 
space-discretization of the 
matrix NLS 
system; 
thus, it can generate  
proper 
space-discrete analogs of 
various multicomponent 
NLS equations~\cite{Mana74,YO2,MMP81,Mak82,KuSk81,ZS74,ForKul83,Svi92} 
obtained as reductions of the matrix NLS system. 

The 
space-discrete 
NLS system 
derived in this paper 
has a rational nonlinearity, so 
it is intrinsically 
more complicated than 
the 
Ablowitz--Ladik 
discretization 
of the NLS system~\cite{AL1,AL76} 
that 
has a polynomial nonlinearity. 
However, 
the rational nonlinearity 
can exhibit 
a saturation 
effect, 
so in some sense it is more 
``physical" 
than the polynomial nonlinearity. 

The space-discrete matrix NLS system appears as 
a member of an infinite hierarchy of commuting flows. 
Other flows 
of 
this integrable 
hierarchy can be obtained 
by 
changing 
the 
temporal 
part 
of the Lax pair.  
In particular, 
we can 
derive a space-discrete analog 
of the matrix mKdV system, which 
admits 
the 
Hermitian conjugation 
(or matrix transposition) 
reduction between the two dependent variables. 
Thus, integrable space-discretizations of 
various multicomponent mKdV 
equations~\cite{YO2,Konop3,Linden2,Atho87,Atho88,Svi93,Svi92} 
can 
be obtained through reductions. 

We 
derive 
the space-discrete matrix NLS hierarchy 
from 
a (slightly generalized) binary 
B\"acklund--Darboux 
transformation 
for the continuous matrix NLS hierarchy
by utilizing it 
as 
the 
underlying 
discrete spectral problem. 
Actually, 
we 
can 
consider 
an arbitrary number of 
B\"acklund--Darboux 
transformations 
%
with 
(generally) 
different values of the 
B\"acklund parameters
and 
assign 
a new discrete independent variable to 
each B\"acklund--Darboux 
transformation; 
that is, 
a single application of 
each 
transformation 
is identified with 
a unit 
shift 
of the corresponding discrete 
independent variable. 
The 
consistency 
of 
such a multidimensional lattice 
is 
guaranteed by 
Bianchi's 
permutability 
theorem
for 
two B\"acklund--Darboux 
transformations on each 
quadrilateral. 
In our approach, 
the dependent variables 
are 
the intermediate 
potentials 
that appear in 
each 
B\"acklund--Darboux transformation, 
so they 
are assigned to the edges 
of the 
lattice~\cite{BoSu08,SuVe03,ABS04,Ve07,PaToVe}.  
Then,  
the permutability
condition for
B\"acklund--Darboux transformations 
on a 
quadrilateral 
can be solved explicitly, 
providing 
a 
%
Yang--Baxter map
(to use 
Veselov's 
terminology~\cite{Ve03,Ve07}); 
this 
is apparently 
related to the 
factorizability property 
of an $N$-soliton collision into pairwise collisions 
in the 
continuous matrix NLS hierarchy (cf.~\cite{Caud2014}). 
The permutability
condition can be regarded as a 
matrix re-factorization problem, 
which 
directly 
provides 
the Lax
representation 
for the Yang--Baxter map~\cite{GoVe04,SuVe03,Ve07,BoSu08}.  
We can identify 
this Yang--Baxter map 
as 
a fully discrete 
analog 
of 
the principal chiral model~\cite{ZaMi78}  
restricted to the space of 
projection matrices.  

This paper is organized as follows. 
In section 2, we introduce 
(a slightly generalized version of) the binary 
B\"acklund--Darboux transformation for the continuous matrix NLS 
hierarchy, 
consider 
it as a discrete 
spectral problem 
and 
associate it with 
a suitable 
isospectral time-evolution equation. 
Then, the compatibility condition 
for 
this Lax pair 
provides 
a lattice system involving some arbitrary parameters; 
with a suitable choice of the parameters, 
it provides 
a proper 
space-discrete analog 
of the matrix NLS system. 
By 
changing 
the 
choice of the 
parameters appropriately, 
we also obtain a proper 
space-discretization of 
the matrix mKdV system.
In section 3, 
we 
solve 
Bianchi's 
permutability 
condition 
for the 
B\"acklund--Darboux transformations 
in such a way that 
a 
Yang--Baxter map of the NLS type
together with 
its 
Lax 
representation 
can be obtained; 
this considerably 
generalizes 
an analogous result of 
Goncharenko and Veselov~\cite{GoVe04,Ve07}. 
Section 4 
is devoted to concluding remarks. 

In appendix~\ref{append}, we 
present 
some result 
obtained by the author in the late 1990s 
but left unpublished. 
We 
consider a Lax-pair representation for 
a vector analog of the sine-Gordon equation~\cite{PR79,EP79}
and describe 
how to discretize one of the two independent variables 
preserving the integrability; 
this is 
somewhat related to 
sections~2 and 3
of 
this 
paper. 
\section
{A proper 
discretization of 
the matrix 
NLS hierarchy}
\label{sect2}

In this section, we 
utilize 
a (slightly generalized) 
binary 
B\"acklund--Darboux transformation 
for the 
continuous matrix NLS 
hierarchy 
as a discrete spectral problem 
and 
derive 
proper 
space-discrete 
analogs of the 
matrix NLS system
and 
the matrix mKdV system.

\subsection
{The continuous 
matrix NLS hierarchy}

In 
1974, Zakharov and Shabat~\cite{ZS74} proposed 
the 
matrix 
generalization
of the 
NLS system~\cite{AKNS73,AKNS74}: 
%
%
\begin{equation} 
\label{mNLS}
\left\{ 
\begin{split}
& \mathrm{i} Q_{t_2} + Q_{xx} - 2Q R Q = O, 
\\[0.5mm]
& \mathrm{i} R_{t_2} - R_{xx} + 2R Q R = O, 
\end{split} 
\right. 
\end{equation}
where 
the dependent variables 
$Q$ and $R$ are \mbox{$l_1 \times l_2$} 
and \mbox{$l_2 \times l_1$} 
(generally rectangular) 
matrices
and 
the subscripts 
denote 
the partial differentiation. 
The symbol
$O$ on the right-hand side of the equations 
is used 
to stress 
that the dependent variables 
can take
values in
matrices. 
The 
matrix NLS system 
(\ref{mNLS}), 
as well as 
related equations, 
has been 
studied 
intensively 
in recent years 
(see, {\it e.g.},~\cite{DM2010}
and references cited therein). 

The matrix NLS system (\ref{mNLS})
is obtained as 
the compatibility condition 
for 
the overdetermined 
linear system 
of 
partial differential equations~\cite{Zakh,Konop1}: 
\begin{align}
& \left[
\begin{array}{c}
 \Psi_1  \\
 \Psi_2 \\
\end{array}
\right]_x 
= \left[
\begin{array}{cc}
-\mathrm{i}\zeta I_{l_1} & Q \\
 R & \mathrm{i}\zeta I_{l_2} \\
\end{array}
\right] 
\left[
\begin{array}{c}
 \Psi_1  \\
 \Psi_2 \\
\end{array}
\right],
\label{NLS-U}
\\[1.5mm]
& \left[
\begin{array}{c}
 \Psi_1  \\
 \Psi_2 \\
\end{array}
\right]_{t_2} 
= \left[
\begin{array}{cc}
-2\mathrm{i}\zeta^2 I_{l_1} -\mathrm{i} QR & 2 \zeta Q + \mathrm{i} Q_x \\
 2 \zeta R - \mathrm{i} R_x & 2\mathrm{i}\zeta^2 I_{l_2} +\mathrm{i} RQ \\
\end{array}
\right]
\left[
\begin{array}{c}
 \Psi_1  \\
 \Psi_2 \\
\end{array}
\right]. 
\label{NLS-V}
\end{align}
%
Here, 
$\zeta$ is 
a constant spectral parameter, 
and 
$I_{l_1}$ and $I_{l_2}$ 
are the \mbox{$l_1 \times l_1 $} and \mbox{$l_2 \times l_2$} 
unit 
matrices, respectively. 
In the following, 
we suppress 
the indices of 
the 
unit matrices,  
noting that the dependent variables can take 
values in 
(generally) rectangular matrices.  
%
%
Equations 
(\ref{NLS-U}) and (\ref{NLS-V}) 
comprise 
the Lax-pair representation~\cite{Lax} 
for 
the matrix NLS system 
(\ref{mNLS}). 
Any constant 
scalar 
matrix 
can be added 
to each 
Lax 
matrix, 
which 
does not affect 
the compatibility condition. 

The matrix NLS system (\ref{mNLS})
is 
a positive 
flow 
in a bi-infinite hierarchy 
of mutually 
commuting flows, 
which 
are all 
associated with 
the 
same 
spectral problem (\ref{NLS-U}). 
The 
next higher flow 
in the matrix NLS hierarchy 
is a matrix 
generalization~\cite{Zakh,Konop1} 
of the complex mKdV 
equation~\cite{AKNS73,AKNS74,Hirota73JMP}, 
which reads 
\begin{equation} 
\label{mmKdV}
\left\{ 
\begin{split}
& Q_{t_3} + Q_{xxx} - 3Q_x R Q -3 QRQ_x = O, 
\\[0.5mm]
& R_{t_3} + R_{xxx} -3 R_x Q R  -3 RQR_x = O. 
\end{split}
\right.
\end{equation}
The first negative flow
in the matrix NLS hierarchy
is a matrix analog of the 
complex sine-Gordon 
equation,
which 
we 
rewrite 
in 
the 
form:\footnote{
The 
first negative flow 
of the 
scalar 
NLS hierarchy~\cite{AKNS73,AKNS74}  
admits 
a number of 
different expressions, 
which are 
related 
through 
simple 
transformations
of dependent variables. 
Such different (but 
essentially 
equivalent)
expressions 
are often 
called by 
different names 
in correspondence 
to 
different
physical phenomena 
(or 
mathematical
objects),  
such as 
%
self-induced transparency (or Maxwell--Bloch) 
equations~\cite{Lamb73PRL,AKN74JMP,AKNS74} 
(also see \S 4.4.b 
of~\cite{AS81}), 
stimulated Raman scattering, 
complex sine-Gordon equation, 
Pohlmeyer--Lund--Regge system, 
reduced nonlinear $\sigma$-model, 
etc.
There are 
too many 
references 
to 
mention 
here.
}
\begin{equation} 
\label{msG}
\left\{ 
\begin{split}
&  Q_{t_{-1}} + 2 u ( I+vu)^{-1} =O, 
\\[0.5mm]
& R_{t_{-1}} - 2 v (I+uv)^{-1} =O, 
\\[0.5mm]
& u_x + 2\mathrm{i} k u = Q - uRu,
\\[0.5mm]
& v_x - 2\mathrm{i} k v = -R + vQv.
\end{split}
\right.
\end{equation}
As long as we consider 
the first negative flow (\ref{msG}) for a single 
fixed 
value of 
$k$, 
the free parameter 
$k$ 
is 
nonessential 
and 
can be 
set equal to zero 
by 
applying 
a 
point transformation 
involving 
$\mathrm{e}^{\pm 2\mathrm{i} k x}$; 
thus, it reduces to 
the simpler form: 
\begin{equation} 
\label{msG2}
\left\{ 
\begin{split}
&  Q_{t_{-1}} + 2 u ( I+vu)^{-1} =O, 
\\[0.5mm]
& R_{t_{-1}} - 2 v (I+uv)^{-1} =O, 
\\[0.5mm]
& u_x = Q - uRu,
\\[0.5mm]
& v_x = -R + vQv.
\end{split}
\right.
\end{equation}
However, 
it is often 
more 
convenient to leave $k$ 
as 
a free 
parameter. 
Note that 
we can consider 
an arbitrary 
linear 
superposition
of 
the first negative flow (\ref{msG}) for 
different values of $k$~\cite{AKN74JMP}. 
In the case of scalar dependent variables, 
we can rewrite 
(\ref{msG}) (or (\ref{msG2})) 
as a closed 
non-evolutionary system for $(Q, R)$ or $(u, v)$.

The 
time part of the Lax-pair representation
(\ref{NLS-V}) for (\ref{mNLS})
is now 
replaced 
with
\begin{align}
& \left[
\begin{array}{c}
 \Psi_1  \\
 \Psi_2 \\
\end{array}
\right]_{t_3} 
= \left[
\begin{array}{cc}
-4\mathrm{i}\zeta^3 I -2\mathrm{i} \zeta QR +Q_x R - QR_x 
 & 4 \zeta^2 Q + 2\mathrm{i}\zeta Q_x -Q_{xx} + 2QRQ \\
 4 \zeta^2 R - 2 \mathrm{i}\zeta R_x -R_{xx} + 2RQR 
 & 4 \mathrm{i}\zeta^3 I + 2\mathrm{i} \zeta RQ +R_x Q - RQ_x\\
\end{array}
\right]
\left[
\begin{array}{c}
 \Psi_1  \\
 \Psi_2 \\
\end{array}
\right] 
\nonumber
\end{align}
for (\ref{mmKdV}) 
and  
\begin{align}
& \left[
\begin{array}{c}
 \Psi_1  \\
 \Psi_2 \\
\end{array}
\right]_{t_{-1}} 
= \frac{\mathrm{i}}{\zeta -k} \uwave{\left\{
\left[
\begin{array}{cc}
 I & \\
 & O \\
\end{array}
\right] + 
\left[
\begin{array}{cc}
 -I & u \\
 v & I \\
\end{array}
\right]^{-1}
\right\} }
\left[
\begin{array}{c}
 \Psi_1  \\
 \Psi_2 \\
\end{array}
\right] 
\label{sG-V}
\end{align} 
for (\ref{msG}),  
respectively. 
Here, 
the underscored part, 
\begin{align}
P &:= \left[
\begin{array}{cc}
 I & \\
  & O \\
\end{array}
\right]
+ \left[
\begin{array}{cc}
 -I & u \\
 v & I \\
\end{array}
\right]^{-1}
\nonumber 
\\[1mm]
& \hphantom{:}= 
\left[
\begin{array}{cc}
 u v \left( I+u v \right)^{-1}
	& u \left( I+v u \right)^{-1} \\
 v \left( I+u v \right)^{-1} & 
	\left( I+v u \right)^{-1}\\
\end{array}
\right]
\nonumber 
\\[1mm]
& \hphantom{:}= 
\left[
\begin{array}{c}
 u \\
 I \\
\end{array}
\right] \left( I+v u \right)^{-1}
\left[
\begin{array}{cc}
 v & I \\
\end{array}
\right]
\nonumber \\[1mm]
& \hphantom{:}= 
\left[
\begin{array}{cc}
 -I & u \\
 v & I \\
\end{array}
\right]
\left[
\begin{array}{cc}
 O & \\
  & I \\
\end{array}
\right]
\left[
\begin{array}{cc}
 -I & u \\
 v & I \\
\end{array}
\right]^{-1},
\label{Puv0}
\end{align}
%
is a projection matrix, 
i.e., 
it 
satisfies
the condition 
\mbox{$P^2 = P$}. 
Note, incidentally, that 
the 
free 
parameter $k$ 
in (\ref{msG}) and (\ref{sG-V}) 
can
be 
generalized to 
an arbitrary function of $t_{-1}$. 

One of the most important properties of the matrix NLS hierarchy, 
which is crucial for 
its 
physical 
applications, 
is that 
it admits 
the Hermitian conjugation 
reduction between the 
pair of 
dependent variables 
$Q$ and $R$~\cite{Mak82} 
(also see~\cite{YO2,Ab78,New79,MMP81,ZakShul82}); 
this reduction 
results in 
a self-focusing, 
self-defocusing 
or 
mixed focusing-defocusing 
nonlinearity,  
which 
is briefly summarized 
in 
\S2.1 of our previous paper~\cite{Tsu13}. 
In particular, 
by imposing the 
reduction 
\mbox{$R = 
- Q^\dagger$} 
on 
the 
matrix NLS system (\ref{mNLS}) 
where the dagger denotes the Hermitian 
conjugation, 
we obtain 
the self-focusing matrix NLS equation~\cite{ZS74}: 
\begin{equation}
\label{mNLSeq}
 \mathrm{i} Q_{t_2} + Q_{xx} + 2Q Q^\dagger Q = O.
\end{equation}
In the same manner, 
the reduction \mbox{$R = - Q^\dagger$} 
simplifies  
(\ref{mmKdV}) 
to the matrix 
complex mKdV equation:
\begin{equation}
\label{mmKdVeq}
 Q_{t_3} + Q_{xxx} + 3Q_x Q^\dagger Q + 3 Q Q^\dagger Q_x = O. 
\end{equation}

Another 
important 
property of the matrix NLS hierarchy 
is that 
the odd-order flows 
admit 
the matrix transposition 
reduction between 
the pair of 
variables 
$Q$ and $R$. 
In particular, 
the matrix mKdV equation~\cite{Linden2,Atho87,Atho88}, 
\begin{equation}
\label{mmKdVeq2}
 Q_{t_3} + Q_{xxx} + 3Q_x Q^T Q + 3 Q Q^T Q_x = O, 
\end{equation}
is obtained by imposing the reduction 
\mbox{$R = - Q^T$} on (\ref{mmKdV}), 
where the superscript ${}^T$ denotes the transpose of a matrix. 
In the case of a vector dependent variable, 
(\ref{mmKdVeq2}) gives 
the vector mKdV equation~\cite{YO2,Konop3}: 
\begin{align}
\label{vmKdV}
& \vt{q}_{t_3} + \vt{q}_{xxx} 
  + 3 \sca{\vt{q}}{\vt{q}_x}\vt{q}
  + 3 \sca{\vt{q}}{\vt{q}}\vt{q}_x 
= \vt{0}.
\end{align}
If $\vt{q}$ is a two-component real-valued vector, i.e., 
\mbox{$\vt{q}= (q_1, q_2)$}, \mbox{$q_1, q_2 \in \mathbb{R}$}, 
the vector mKdV equation (\ref{vmKdV}) can be rewritten as 
a 
single
complex-valued mKdV equation 
for 
\mbox{$q := q_1 + \mathrm{i} q_2 
$}; 
this 
is often referred to as the Sasa--Satsuma equation~\cite{Sasa91}.

Any proper discretization of the matrix NLS hierarchy 
should 
retain 
the feasibility 
of 
such reductions. 
In short, 
a discrete analog 
of 
the 
matrix NLS 
hierarchy 
is physically 
meaningful 
only if 
it is integrable and 
admits 
the Hermitian conjugation 
(or matrix 
transposition) 
reduction 
similar 
to 
the continuous case. 
That is, 
it should be able to provide integrable 
discretizations for 
the reduced equations such as (\ref{mNLSeq})--(\ref{vmKdV}). 

%
%
%
%
%
%
%
%
%

\subsection{B\"acklund--Darboux transformation 
as a discrete spectral problem}

To obtain 
a proper 
discrete analog of the matrix NLS hierarchy, 
we 
start with 
a slightly generalized 
version 
of 
the binary B\"acklund--Darboux transformation~\cite{Chud2,Sall82,ZS79} 
for the continuous matrix NLS hierarchy 
as 
given by 
\begin{align}
& \left[
\begin{array}{c}
 \widetilde{\Psi}_1  \\
 \widetilde{\Psi}_2 \\
\end{array}
\right]
= \left\{ \left[
\begin{array}{cc}
(  \zeta \delta + \alpha ) I &  \\
  & (- \zeta \gamma + \beta) I \\
\end{array}
\right] - (\alpha \gamma + \beta \delta)
 \left[
\begin{array}{cc}
\gamma I & u \\
 v & \delta I \\
\end{array}
\right]^{-1}
\right\}
\left[
\begin{array}{c}
 \Psi_1  \\
 \Psi_2 \\
\end{array}
\right]. 
\label{gDB1}
\end{align}
Here, $\alpha$, $\beta$, $\gamma$ and $\delta$ are arbitrary parameters, 
except that 
they are required 
to satisfy 
the 
condition 
\mbox{$\alpha \gamma + \beta \delta \neq 0$}; 
the special case \mbox{$\delta = -\gamma \hspace{1pt}(\neq 0)$} corresponds to 
the conventional binary B\"acklund--Darboux transformation~\cite{Sall82} 
or the Zakharov--Shabat dressing 
method~\cite{Chud2,ZS79}, up to an 
inessential 
overall factor. 
Unlike the 
usual 
formulation of the binary 
B\"acklund--Darboux transformation, 
we 
do 
not express 
the intermediate 
potentials $u$ and $v$
in terms of 
linear eigenfunctions of the original 
Lax-pair representation~\cite{Chud1,Chud2,Kono82}. 
Alternatively, 
we will 
consider $u$ and $v$ 
as a pair of new dependent variables 
defined on a lattice, 
wherein 
the 
lattice index 
\mbox{$n \in \mathbb{Z}$}
is understood intuitively 
as 
the number 
of iterations of the 
B\"acklund--Darboux transformation. 
%
To make 
it 
more explicit, we 
rewrite (\ref{gDB1}) 
as a discrete spectral problem
\begin{equation}
\left[
\begin{array}{c}
 \Psi_{1, n+1}  \\
 \Psi_{2, n+1} \\
\end{array}
\right]
= L_n 
\left[
\begin{array}{c}
 \Psi_{1,n}  \\
 \Psi_{2,n} \\
\end{array}
\right], 
\label{gDB2}
\end{equation}
%
where the 
Lax matrix $L_n$ 
is given 
by 
\begin{align}
L_n & =  \left[
\begin{array}{cc}
(  \zeta \delta + \alpha ) I &  \\
  & (- \zeta \gamma + \beta) I \\
\end{array}
\right] - (\alpha \gamma + \beta \delta)
 \left[
\begin{array}{cc}
\gamma I & u_n \\
 v_n & \delta I \\
\end{array}
\right]^{-1}
\nonumber \\[2mm]
& =  \left[
\begin{array}{cc}
(  \zeta \delta + \alpha ) I &  \\
  & (- \zeta \gamma + \beta) I \\
\end{array}
\right] 
\nonumber \\[1mm]
& \hphantom{=} \;\, \mbox{}
- (\alpha \gamma + \beta \delta)
 \left[
\begin{array}{cc}
\delta (\gamma \delta I - u_n v_n)^{-1}  
 & -(\gamma \delta I - u_n v_n)^{-1}u_n \\
 -(\gamma \delta I - v_n u_n)^{-1}v_n 
 & \gamma (\gamma \delta I - v_n u_n)^{-1}  \\
\end{array}
\right]
\nonumber \\[2mm]
&= \left[
\begin{array}{cc}
\delta I & -u_n \\
 -v_n &  \gamma I \\
\end{array}
\right]
\left[
\begin{array}{cc}
(\zeta \gamma - \beta)  I &  \\
  &  -(\zeta \delta + \alpha ) I \\
\end{array}
\right] 
 \left[
\begin{array}{cc}
\gamma I & u_n \\
 v_n & \delta I \\
\end{array}
\right]^{-1}
\nonumber \\[2mm]
&= \left[
\begin{array}{cc}
\gamma I & u_n \\
 v_n & \delta I \\
\end{array}
\right]^{-1}
\left[
\begin{array}{cc}
(\zeta \gamma - \beta)  I &  \\
  &  -(\zeta \delta + \alpha) I \\
\end{array}
\right] 
\left[
\begin{array}{cc}
\delta I & -u_n \\
 -v_n &  \gamma I \\
\end{array}
\right]. 
\nonumber \\ & 
\label{Lax-Ln1}
\end{align}
Actually, 
the 
B\"acklund 
parameters 
$\alpha$, $\beta$, $\gamma$ and $\delta$
satisfying 
the condition 
\mbox{$\alpha \gamma + \beta \delta \neq 0$}
can 
be 
arbitrary functions of 
the 
discrete independent variable $n$, 
but for 
brevity we usually 
consider 
them as constants. 
Each 
of the 
four 
expressions 
in (\ref{Lax-Ln1}) 
has its own advantages. 
%

\subsection{Isospectral time evolution}
\label{sec2.3}

To 
compose 
a Lax pair, 
we need to associate 
(\ref{gDB2}) with 
a suitable 
isospectral 
time-evolution equation, 
\begin{equation}
\left[
\begin{array}{c}
 \Psi_{1, n}  \\
 \Psi_{2, n} \\
\end{array}
\right]_t
= M_n 
\left[
\begin{array}{c}
 \Psi_{1,n}  \\
 \Psi_{2,n} \\
\end{array}
\right]. 
\label{general_M}
\end{equation}
Here, 
the spectral parameter $\zeta$ 
involved in $L_n$ and $M_n$ 
is isospectral, 
i.e., 
\mbox{$\zeta_t=0$}. 
%
The compatibility condition 
for 
(\ref{gDB2}) and (\ref{general_M})
is given by (a space-discrete 
version 
of) the zero-curvature 
equation~\cite{AL1,Kako,AL76,
Ize81}
\begin{equation}
 L_{n,t} +L_n M_n - M_{n+1}L_n = O, 
\label{Lax_eq}
\end{equation}
where
$L_n$ and $M_n$  
comprise 
the 
Lax pair. 
For the Lax matrix $L_n$ 
in (\ref{Lax-Ln1}), 
we 
use 
the first expression 
to compute $L_{n,t}$ 
with the aid of 
the formula 
\mbox{$(X^{-1})_t = - X^{-1} X_t X^{-1}$} 
valid for a square matrix $X$
so that 
the 
equations of motion 
for $u_{n}$ and $v_{n}$
can be obtained 
explicitly. 

Let us first consider 
the original continuous spectral problem 
(\ref{NLS-U}), which now reads 
\begin{align}
\left[
\begin{array}{c}
 \Psi_{1,n}  \\
 \Psi_{2,n} \\
\end{array}
\right]_x 
= \left[
\begin{array}{cc}
-\mathrm{i}\zeta I & Q_n \\
 R_n & \mathrm{i}\zeta I \\
\end{array}
\right] 
\left[
\begin{array}{c}
 \Psi_{1,n}  \\
 \Psi_{2,n} \\
\end{array}
\right]. 
\label{conNLS}
\end{align}
%
The B\"acklund--Darboux transformation (\ref{gDB2}) with (\ref{Lax-Ln1}) 
preserves the form of 
the spectral problem (\ref{conNLS}) 
invariant. 
In the general case where the B\"acklund parameters 
$\alpha$, $\beta$, $\gamma$ and $\delta$ 
depend on $n$ (\mbox{$\alpha \to\alpha_n$
etc.}), 
the 
compatibility condition 
(i.e., (\ref{Lax_eq}) with 
$t$ replaced by $x$)
provides a system of differential-difference equations: 
\begin{equation} 
\label{dmsG}
\left\{ 
\begin{split}
& \gamma_n Q_{n+1} + \delta_n Q_n + 2 \mathrm{i} (\alpha_n \gamma_n 
	+\beta_n\delta_n )u_n (\gamma_n \delta_n I-v_n u_n )^{-1} =O, 
\\[0.5mm]
& \delta_n R_{n+1} + \gamma_n R_n + 2 \mathrm{i} (\alpha_n \gamma_n 
	+\beta_n\delta_n ) v_n (\gamma_n \delta_n I-u_n v_n )^{-1} =O, 
\\[0.5mm]
& (\alpha_n \gamma_n +\beta_n\delta_n ) u_{n,x} + \alpha_n \gamma_n^2 Q_{n+1} 
	- \beta_n \delta_n^2 Q_n - u_n (\alpha_n R_{n+1} - \beta_n R_n) u_n = O,
\\[0.5mm]
& (\alpha_n \gamma_n +\beta_n\delta_n ) v_{n,x} +\beta_n \delta_n^2 R_{n+1}
	-\alpha_n \gamma_n^2 R_n - v_n (\beta_n Q_{n+1} -\alpha_n Q_n) v_n = O.
\end{split}
\right.
\end{equation}
%
When \mbox{$\gamma_n \delta_n \neq 0$}, 
the first two 
equations 
relate the 
new potentials $Q_{n+1}$ and $R_{n+1}$ 
to the old potentials $Q_{n}$ and $R_{n}$ 
in the spectral problem (\ref{conNLS}) 
through 
the intermediate potentials $u_n$ and $v_n$ 
(cf.~\cite{Kono82}); 
in this case, 
(\ref{dmsG}) can be reformulated 
as 
Riccati equations for $u_n$ and $v_n$: 
\begin{equation} 
\label{uv_exp}
\left\{ 
\begin{split}
& \delta_n u_{n,x} = 
2 \mathrm{i} \alpha_n u_n + \delta_n^2 Q_n 
	- u_n R_n u_n, 
\\[0.5mm]
& \gamma_n v_{n,x}  = 
2 \mathrm{i} \beta_n v_{n}
	+ \gamma_n^2 R_n - v_n Q_n v_n, 
\\[0.5mm]
& -\gamma_n u_{n,x} =
2 \mathrm{i} \beta_n u_n + \gamma_n^2 Q_{n+1} 
	- u_n R_{n+1} u_n, 
\\[0.5mm]
& -\delta_n v_{n,x} = 
2 \mathrm{i} \alpha_n v_{n}
	+ \delta_n^2 R_{n+1} - v_n Q_{n+1} v_n. 
\end{split}
\right.
\end{equation}
Thus, 
the intermediate potentials 
$u_n$ and $v_n$ can be expressed 
in terms of 
the 
linear eigenfunctions of the 
spectral problem (\ref{conNLS})~\cite{Chud1,Kono82}, {\it e.g.} 
\[
\left. v_n= \gamma_n 
	\Psi_{2,n} \Psi_{1,n}^{-1}\right|_{\zeta=\frac{\beta_n}{\gamma_n}}
\left. = -\delta_n 
	\Psi_{2,n+1} \Psi_{1,n+1}^{-1}\right|_{\zeta=-\frac{\alpha_n}{\delta_n}}, 
\]
where 
$\Psi_{1,n}$ and $\Psi_{1,n+1}$
are square 
matrices. 
In the case of scalar dependent variables, 
we can eliminate $Q_n$ and $R_n$ 
from (\ref{uv_exp})  
to obtain a closed 
system for $u_n$ and $v_n$.  

With a suitable choice of the parameters, 
(\ref{dmsG}) can be considered as 
an integrable 
time-discretization
of 
the matrix sine-Gordon system (\ref{msG2}). 
Indeed, 
in the case of \mbox{$\beta_n = - \alpha_n$} and 
\mbox{$\delta_n = - \gamma_n$}, (\ref{dmsG}) 
provides 
a (generally nonuniform) 
discrete-time analog of (\ref{msG2}):
\begin{equation} 
\label{dmsG2}
\left\{ 
\begin{split}
& Q_{n+1} - Q_n - 4 \mathrm{i} \alpha_n 
	u_n (\gamma_n^2 I + v_n u_n )^{-1} =O, 
\\[0.5mm]
& R_{n+1} - R_n + 4 \mathrm{i} \alpha_n 
	v_n (\gamma_n^2 I+u_n v_n )^{-1} =O, 
\\[0.5mm]
& 2 \gamma_n u_{n,x} + \gamma_n^2 (Q_{n+1} +Q_n) - u_n (R_{n+1} + R_n) u_n = O,
\\[0.5mm]
& 2 \gamma_n v_{n,x} - \gamma_n^2 ( R_{n+1} + R_n) 
	+ v_n (Q_{n+1} + Q_n) v_n = O.
\end{split}
\right.
\end{equation}
%

To obtain 
proper space-discretizations 
of 
positive flows of the matrix NLS hierarchy, 
we 
consider a temporal Lax matrix $M_n$
that 
matches well 
with 
the factorized form of the spatial 
Lax matrix $L_n$ 
in 
(\ref{Lax-Ln1}). 
Note
that the first expression in (\ref{Lax-Ln1}) implies 
the $\zeta$-independence of 
$L_{n,t}$, 
so 
\mbox{$L_n M_n - M_{n+1}L_n$} in (\ref{Lax_eq}) 
is 
also
$\zeta$-independent. 
A 
simple 
ansatz for $M_n$
satisfying 
this condition automatically
is 
\begin{align}
& M_n 
= 
\left[
\begin{array}{cc}
\gamma I & u_n \\
 v_n & \delta I \\
\end{array}
\right]
\left[
\begin{array}{cc}
\frac{1}{\zeta \gamma -\beta} F_n & \\
  & \frac{1}{\zeta \delta + \alpha} G_n \\
\end{array}
\right] 
\left[
\begin{array}{cc}
\gamma I & u_{n-1} \\
 v_{n-1} & \delta I \\
\end{array}
\right] 
+ 
\left[
\begin{array}{cc}
c I & \\
 & d I \\
\end{array}
\right],  
\label{Lax-Mn1}
\end{align}
where $F_n$ and $G_n$ 
are $\zeta$-independent 
square matrices
and $c$ and $d$ are $\zeta$-independent 
parameters. 
%
The $\zeta$-dependence of 
the main part of $M_n$ 
in (\ref{Lax-Mn1}) 
implies that it 
can be considered as 
a linear superposition of 
the temporal Lax matrix in 
(\ref{sG-V}) at 
two 
particular
values of $k$ (up to the addition of a constant scalar matrix), 
i.e., it corresponds to the two 
``first" negative 
flows of the matrix NLS hierarchy. 
Substituting (\ref{Lax-Ln1}) and (\ref{Lax-Mn1}) 
into (\ref{Lax_eq}), 
we 
obtain 
recurrence relations for 
$F_n$ and $G_n$; 
to satisfy them identically, 
we 
set  
\begin{align}
& F_n = a \left( \alpha \gamma +\beta \delta \right) 
\left( \gamma^2 I + u_{n-1} v_n \right)^{-1}, 
\hspace{5mm}
G_n = b \left( \alpha \gamma +\beta \delta \right) 
\left( \delta^2 I + v_{n-1} u_n \right)^{-1}, 
\nonumber
\end{align}
where 
$a$ and $b$ are arbitrary constants. 
In fact, 
$a$ and $b$, 
as well as $c$ and $d$ in (\ref{Lax-Mn1}), 
can depend arbitrarily on 
the time variable 
$t$, but 
we do not 
consider 
it 
in this paper. 
With the above choice of $F_n$ and $G_n$, 
(\ref{Lax_eq}) 
for 
(\ref{Lax-Ln1}) and (\ref{Lax-Mn1}) 
provides 
an evolutionary system of differential-difference equations: 
\begin{equation} 
\label{sdNLS}
\left\{ 
\begin{split}
& u_{n,t}  + a \left( \gamma \delta I - u_{n} v_n \right)
	\left( \gamma^2 I + u_{n-1} v_n \right)^{-1} 
	\left( \gamma u_n + \delta u_{n-1} \right)
\\ 
& 
\mbox{} + b \left( \gamma u_{n+1} + \delta u_{n} \right)
 \left( \delta^2 I + v_{n} u_{n+1} \right)^{-1} 
 \left( \gamma \delta I - v_{n} u_n \right)
 - (c-d) u_n = O, 
\\[1.5mm]
& v_{n,t}  - b \left( \gamma \delta I - v_{n} u_n \right)
	\left( \delta^2 I + v_{n-1} u_n \right)^{-1} 
	\left( \delta v_n + \gamma v_{n-1} \right)
\\ 
& \mbox{} - a \left( \delta v_{n+1} + \gamma v_n \right) 
 \left( \gamma^2 I + u_{n} v_{n+1} \right)^{-1} 
 \left( \gamma \delta I - u_{n} v_n \right)
 + (c-d) v_n = O, 
\end{split}
\right.
\end{equation}
which is independent of $\alpha$ and $\beta$. 
Note that 
the order of 
products in 
(\ref{sdNLS}) 
can be changed
using 
identities such as 
\begin{align}
& \left( \gamma \delta I - u_{n} v_n \right)
	\left( \gamma^2 I + u_{n-1} v_n \right)^{-1} 
	\left( \gamma u_n + \delta u_{n-1} \right)
\nonumber \\
&= \delta u_n + \frac{\delta^2}{\gamma} u_{n-1} 
-\frac{1}{\gamma} \left( \gamma u_n + \delta u_{n-1} \right) v_n
 \left( \gamma^2 I + u_{n-1} v_n \right)^{-1} 
 \left( \gamma u_n + \delta u_{n-1} \right)
\nonumber \\
&= \left( \gamma u_n + \delta u_{n-1} \right)
	\left( \gamma^2 I + v_n u_{n-1} \right)^{-1} 
	\left( \gamma \delta I - v_n u_{n}  \right). 
\nonumber 
\end{align}

In the case of scalar 
$u_n$ and $v_n$, 
(\ref{sdNLS}) 
is related 
to 
known integrable systems 
such as the lattice Heisenberg ferromagnet model~\cite{Ishi82,Skl82,Fad84,Hal82}  
through simple 
changes of dependent variables 
(also see (2.19)--(2.20) in~\cite{Veks11}
together with~\cite{AdSha06}); 
thus, 
it is 
not really a new integrable system. 
However, 
(\ref{sdNLS}) in the general matrix case, 
its reductions and 
identification as a discrete analog of 
the matrix NLS (or mKdV) 
equation 
have not been 
reported 
in the literature. 

Considering 
its relation to 
the space-discrete Kaup--Newell system 
(see Propositions~2.1
and 2.2 in~\cite{Me2012} 
with 
a variable change \mbox{$v_n \to v_n^{-1}$}), 
we can show 
that 
the lattice system 
(\ref{sdNLS}) possesses 
an ultralocal 
(but 
noncanonical) 
Poisson 
bracket. 
For scalar $u_n$ and $v_n$, 
it 
can be 
written 
as 
\mbox{$u_{n,t} = \{ u_n, H \}$} and \mbox{$v_{n,t} = \{ v_n, H \}$}
with the Hamiltonian and the Poisson bracket 
given 
by 
\[
H = 
\sum_n \left[ a \log \left( \frac{ \gamma^2 + u_{n-1} v_{n}}
	{\gamma \delta - u_{n} v_{n}} \right)
 + b \log \left( \frac{ \delta^2 + u_{n+1} v_{n}}
	{\gamma \delta - u_{n} v_{n}} \right) 
 + 
\frac{c-d}{\gamma \delta - u_{n} v_{n}}
\right]
\]
and 
\[
\{ u_m, u_n \} = \{ v_m, v_n \} =0, \hspace{5mm} 
\{ u_m, v_n \} = \delta_{m n} \left( \gamma \delta - u_{n} v_{n} \right)^2, 
\]
respectively. Here, $\delta_{m n}$ is 
the Kronecker delta, which 
has nothing to do with the free parameter $\delta$. 

By rescaling the 
variables and 
parameters 
as 
\mbox{$u_n \to \gamma u_n$}, \mbox{$v_n \to \gamma v_n$}, 
\mbox{$a \to \gamma a $},  \mbox{$b \to 
b/\gamma^2$} and setting \mbox{$c = b/\gamma$}, 
we can rewrite 
(\ref{sdNLS}) 
as 
\begin{equation} 
\label{sdNLS2}
\left\{ 
\begin{split}
& u_{n,t}  + a \left( \delta I - \gamma u_{n} v_n \right)
	\left( I + u_{n-1} v_n \right)^{-1} 
	\left( \gamma u_n + \delta u_{n-1} \right)
\\ 
& 
\mbox{} + \frac{b}{\gamma} \left( \gamma u_{n+1} + \delta u_{n} \right)
 \left( \delta^2 I + \gamma^2 v_{n} u_{n+1} \right)^{-1} 
 \left( \delta I - \gamma v_{n} u_n \right)
 -  \frac{b}{\gamma} u_n + d u_n = O, 
\\[1.5mm]
& v_{n,t}  - \frac{b}{\gamma} \left(  \delta I - \gamma v_{n} u_n \right)
	\left( \delta^2 I + \gamma^2 v_{n-1} u_n \right)^{-1} 
	\left( \gamma v_{n-1} + \delta v_n \right)
\\ 
& \mbox{} - a \left( \gamma v_n + \delta v_{n+1} \right) 
 \left( I + u_{n} v_{n+1} \right)^{-1} 
 \left(  \delta I - \gamma u_{n} v_n \right)
 +  \frac{b}{\gamma} v_n -d v_n = O. 
\end{split}
\right.
\end{equation}
Then, by taking the limit 
\mbox{$\gamma \to 0$}, 
(\ref{sdNLS2}) reduces 
to a known 
integrable 
space-discretization of the NLS system~\cite{GI82,Tu}: 
\begin{equation} 
\nonumber 
\left\{ 
\begin{split}
& u_{n,t}  + a \delta^2 \left( I + u_{n-1} v_n \right)^{-1} u_{n-1}
  + \frac{b}{\delta} \left( u_{n+1} - u_{n} v_{n} u_n \right)
 + d u_n = O, 
\\
& v_{n,t}  - \frac{b}{\delta} \left( v_{n-1} - v_{n} u_n v_n \right) 
  - a \delta^2 v_{n+1} \left( I + u_{n} v_{n+1} \right)^{-1} 
 -d v_n = O,
\end{split}
\right.
\end{equation}
which 
is associated with an elementary auto-B\"acklund
transformation for the continuous 
matrix 
NLS hierarchy~\cite{Kono82,Chud1,Chud2} 
as 
outlined in~\cite{Comment}.
This 
limiting case 
no longer 
admits the complex
(or Hermitian) conjugation reduction between 
$u_n$ and $v_n$, 
so it 
does not 
provide 
a physically meaningful 
discretization of the NLS 
system. 

When the 
parameters 
satisfy 
the conditions 
\mbox{$\delta = \gamma^\ast$}, 
\mbox{$b=-a^\ast$} and \mbox{$d=c^\ast$}, 
the lattice system 
(\ref{sdNLS}) 
admits 
not only the complex conjugation reduction 
for square matrices $u_n$ and $v_n$ 
but also the Hermitian conjugation reduction 
for (generally) rectangular matrices $u_n$ and $v_n$. 
In particular, 
in the case of 
\mbox{$\delta = \gamma = 1$}, 
\mbox{$b=-a^\ast$} and \mbox{$d=c^\ast$},
the Hermitian conjugation reduction 
\mbox{$v_n = -u_n^\dagger$} 
simplifies (\ref{sdNLS}) 
to 
\begin{align} 
 u_{n,t} & -a^\ast  \left( I +u_n  u_{n}^\dagger \right)
 \left( I - u_{n+1} u_{n}^\dagger  \right)^{-1} 
 \left( u_{n+1} + u_{n} \right)
\nonumber 
\\ 
& 
\mbox{} + a \left( I + u_{n} u_n^\dagger \right)
	\left( I - u_{n-1} u_n^\dagger \right)^{-1} 
	\left( u_n + u_{n-1} \right)
 - (c-c^\ast) u_n = O. 
\label{sdNLS3}
\end{align}
By further setting 
\mbox{$a =-\mathrm{i}$} and  \mbox{$c =-2\mathrm{i}$}, 
(\ref{sdNLS3}) provides 
a new 
integrable space-discretization of the matrix NLS equation 
(\ref{mNLSeq}), 
\begin{align} 
 \mathrm{i} u_{n,t} & + 
\left( I + u_{n} u_n^\dagger \right) 
 \left( I - u_{n+1} u_{n} ^\dagger \right)^{-1} 
  \left( u_{n+1} + u_{n} \right) 
\nonumber \\ 
& 
\mbox{} 
+ \left( I + u_{n} u_n^\dagger \right)
\left( I - u_{n-1} u_n^\dagger \right)^{-1} 
	\left( u_n + u_{n-1} \right)
  - 4 u_n = O, 
\nonumber
\end{align}
which can 
also 
be written as 
\begin{align} 
 \mathrm{i} u_{n,t} & + \left( u_{n+1} + u_{n-1} - 2 u_n \right) 
+ \left( u_{n+1} + u_{n} \right) u_n^\dagger 
  \left( I - u_{n+1} u_{n}^\dagger \right)^{-1} 
  \left( u_{n+1} + u_{n} \right) 
\nonumber \\ 
& \mbox{} + \left( u_n + u_{n-1} \right) u_n^\dagger 
\left( I - u_{n-1} u_n^\dagger \right)^{-1} 
	\left( u_n + u_{n-1} \right) 
 = O. 
\label{sdNLS5}
\end{align}
This 
equation 
is defined on the three lattice 
sites \mbox{$n-1$}, $n$, \mbox{$n+1$}, so it is 
simpler 
and appears to be more interesting 
than 
the space-discrete matrix NLS equation 
proposed in our previous paper~\cite{Tsuchi02} (see (4.1) therein), 
which depends on five lattice sites 
after imposing the Hermitian conjugation reduction. 
Considering 
further reductions 
of the matrix dependent variable $u_n$, 
we can obtain integrable 
space-discretizations 
of various multicomponent NLS equations~\cite{KuSk81,ForKul83,Svi92}. 
In particular, when $u_n$ is a row (or column) vector, 
(\ref{sdNLS5}) provides a new integrable space-discretization 
of the vector NLS equation 
often referred to as the Manakov model~\cite{Mana74}. 
Equation (\ref{sdNLS5}) corresponds to the self-focusing case, 
but it 
is also possible to consider the self-defocusing case 
or a mixed focusing-defocusing case as 
in the continuous case~\cite{Mak82,YO2,Ab78,New79,MMP81,ZakShul82}. 

Note 
that 
a fully discrete matrix NLS 
equation was 
proposed 
by van der Linden, Nijhoff, Capel and Quispel 
in 1986 
(see (4.17) in~\cite{Linden2}). 
However, it involves a square-root function of 
a general 
non-diagonal matrix 
defined 
using the Taylor series expansion, 
so it 
is not 
an explicit 
and 
closed 
expression. 
In fact, 
even in the scalar case, 
their 
fully 
discrete NLS equation 
(see 
(7.18) in~\cite{QNCL})  
is an implicit equation, which 
does not define the time evolution uniquely. 

In the case of \mbox{$a =-1$} and  \mbox{$c =0$}, 
(\ref{sdNLS3}) provides 
an 
integrable space-discretization of the matrix complex mKdV equation 
(\ref{mmKdVeq})
given by 
\begin{align} 
u_{n,t} & +  \left( I +u_n  u_{n}^\dagger \right)
 \left( I - u_{n+1} u_{n}^\dagger  \right)^{-1} 
 \left( u_{n+1} + u_{n} \right)
\nonumber 
\\ 
& 
\mbox{} 
- 
\left( I + u_{n} u_n^\dagger \right)
	\left( I - u_{n-1} u_n^\dagger \right)^{-1} 
	\left( u_n + u_{n-1} \right) 
= O. 
\nonumber 
\end{align}

When the 
parameters 
satisfy 
the conditions 
\mbox{$\delta = \gamma$}, 
\mbox{$b=-a$} and \mbox{$c=d=0$}, 
the lattice system 
(\ref{sdNLS}) 
admits the matrix transposition 
reduction 
between $u_n$ and $v_n$. 
In particular, by setting 
\mbox{$\delta = \gamma=1$}, 
\mbox{$b=-a=1$}, 
\mbox{$c=d=0$}
and 
\mbox{$v_n = -u_n^T$}, 
(\ref{sdNLS}) reduces to 
an integrable space-discretization of the matrix mKdV 
equation (\ref{mmKdVeq2}):
\begin{align} 
u_{n,t} & +  \left( I +u_n  u_{n}^T \right)
 \left( I - u_{n+1} u_{n}^T  \right)^{-1} 
 \left( u_{n+1} + u_{n} \right)
\nonumber 
\\ 
& 
\mbox{} - \left( I + u_{n} u_n^T \right)
	\left( I - u_{n-1} u_n^T \right)^{-1} 
	\left( u_n + u_{n-1} \right) = O. 
\nonumber 
\end{align}
In the vector case, 
this reads 
\begin{align}
\label{dvmKdV}
\vt{u}_{n,t} 
& + \left( 1+ \sca{\vt{u}_n}{\vt{u}_n} \right)  \left(
\frac{\vt{u}_{n+1} + \vt{u}_n}{1-\sca{\vt{u}_{n+1}}{\vt{u}_n}} 
 - \frac{\vt{u}_{n} + \vt{u}_{n-1}}{1-\sca{\vt{u}_n}{\vt{u}_{n-1}}} \right) 
%
= \vt{0}, 
\end{align}
which provides 
an 
integrable 
space-discrete analog of 
the vector mKdV equation (\ref{vmKdV}). 
In the single-component (i.e., scalar) 
case, this 
can be rewritten in 
a 
more 
familiar form (cf.~(3.5) in~\cite{Hiro73}) 
by setting 
\mbox{$u_n =: \tan w_n$}. 
When 
$\vt{u}_n$ is a real 
two-component vector, i.e., 
\mbox{$\vt{u}_n = (u_n^{(1)}, u_n^{(2)})$}, 
\mbox{$u_n^{(1)}, u_n^{(2)} \in \mathbb{R}$}, 
we can introduce a 
complex-valued function 
\mbox{$\psi_n := u_n^{(1)} + \mathrm{i} u_n^{(2)}$} 
to rewrite 
(\ref{dvmKdV}) 
as 
a single 
evolution 
equation 
for 
$\psi_n$. 
Then, we obtain 
an integrable 
space-discretization of 
the 
complex mKdV equation
often 
referred to as 
the Sasa--Satsuma equation~\cite{Sasa91}. 

When 
\mbox{$\delta = \pm \gamma$}, 
\mbox{$b=-a$} and \mbox{$c=d=0$}, 
we can reduce 
the lattice system 
(\ref{sdNLS}) 
to a single closed equation for $u_n$ 
by setting 
$v_n$ as a constant scalar matrix. 
In particular, by setting 
\mbox{$\delta = \gamma$}, 
\mbox{$b=-a$}, \mbox{$c=d=0$},
\mbox{$v_n=-\gamma^2 I$} and \mbox{$a \gamma=-1$}, 
we obtain 
a 
proper space-discretization of 
the matrix KdV equation~\cite{Lax} (cf.~(\ref{mmKdV}) 
with 
\mbox{$R=\mathrm{const.}$}), which reads 
\begin{align} 
u_{n,t} & + \left( u_{n+1}-u_{n-1} \right) 
 + \left( u_{n+1} + u_{n} \right)
 \left( I - u_{n+1} \right)^{-1} 
 \left( u_{n+1} + u_{n} \right)
\nonumber 
\\ 
& 
\mbox{} - \left( u_{n} + u_{n-1} \right)
	\left( I - u_{n-1} \right)^{-1} 
	\left( u_n + u_{n-1} \right) = O. 
\nonumber 
\end{align}
Alternatively, 
by setting \mbox{$\delta = -\gamma$}, 
\mbox{$b=-a$}, \mbox{$c=d=0$},
\mbox{$v_n=\gamma^2 I$} and \mbox{$a \gamma=1$}, 
we obtain 
\begin{align} 
u_{n,t} & + \left( u_{n+1} + u_{n-1} - 2 u_n \right) 
 - \left( u_{n+1} - u_{n} \right) 
  \left( I + u_{n+1} \right)^{-1} 
  \left( u_{n+1} - u_{n} \right) 
\nonumber \\ 
& \mbox{} - \left( u_n - u_{n-1} \right)
\left( I + u_{n-1} \right)^{-1} 
	\left( u_n - u_{n-1} \right) 
 = O, 
\nonumber 
\end{align}
which is an integrable space-discretization of 
the 
equation
\[
u_t + u_{xx}  -2 u_x (I+u)^{-1} u_x =O
\]
that can be linearized as 
\[
\left[ \left( I+u \right)^{-1} \right]_t 
	+ \left[ \left( I+u \right)^{-1} \right]_{xx} =O. 
\]

\section{Yang--Baxter map} 

In this section, we 
utilize 
the permutability property 
of 
B\"acklund--Darboux transformations 
to 
construct 
a 
Yang--Baxter map, 
which 
provides 
a fully discrete analog 
of the principal chiral model for 
projection matrices 
and admits the Hermitian conjugation reduction.


Let us 
consider 
two copies of 
the 
``same" 
B\"acklund--Darboux transformation 
with different sets of 
B\"acklund parameters 
and intermediate potentials
and assign them  
to 
two 
lattice directions 
$m$ and $n$, respectively. 
A forward shift 
in the $n$-direction 
is defined 
as in (\ref{gDB2}) with (\ref{Lax-Ln1}), 
i.e.\ 
\begin{equation}
\Psi_{m,n+1}
= L_{m,n} 
\Psi_{m,n}
\label{gDB3}
\end{equation}
%
with the 
Lax matrix $L_{m,n}$ 
given 
by 
\begin{align}
L_{m,n} & =  \left[
\begin{array}{cc}
(  \zeta \delta + \alpha ) I &  \\
  & (- \zeta \gamma + \beta) I \\
\end{array}
\right] - (\alpha \gamma + \beta \delta)
 \left[
\begin{array}{cc}
\gamma I & u_{m,n} \\
 v_{m,n} & \delta I \\
\end{array}
\right]^{-1}. 
\label{Lax-Ln2}
\end{align}
Here, 
the condition 
\mbox{$\alpha \gamma + \beta \delta \neq 0$} 
is assumed; 
for brevity, 
the $n$-dependence of the 
B\"acklund 
parameters 
$\alpha$, $\beta$, $\gamma$ and $\delta$ 
is suppressed. 
We are only interested in 
the general 
case 
of \mbox{$\gamma \delta \neq 0$} and do not consider 
the 
limiting 
case 
of 
\mbox{$\gamma \, (\mathrm{or} \; \delta) \to0$} 
where the binary B\"acklund--Darboux transformation 
can reduce 
to an elementary B\"acklund--Darboux transformation. 

Similarly, 
we define 
a forward shift 
in the $m$-direction 
as 
\begin{equation}
\Psi_{m+1,n}
= V_{m,n} \Psi_{m,n}
\label{gDB4}
\end{equation}
with the 
Lax matrix $V_{m,n}$ 
given 
by 
\begin{align}
V_{m,n} & =  \left[
\begin{array}{cc}
(  \zeta d + a ) I &  \\
  & (- \zeta c + b) I \\
\end{array}
\right] - (a c + b d)
 \left[
\begin{array}{cc}
 c I & q_{m,n} \\
 r_{m,n} & d I \\
\end{array}
\right]^{-1}. 
\label{Lax-Ln3}
\end{align}
Note that 
the 
parameters $a$, $b$, $c$ and $d$ 
used 
in this section 
have no direct relationship with
those 
used in subsection~\ref{sec2.3}. 
The condition 
\mbox{$ac+bd \neq 0$} 
is assumed and the $m$-dependence of $a$, $b$, $c$ and $d$ 
is suppressed. 
In addition, 
we assume \mbox{$cd \neq 0$} and do not consider 
the 
limiting 
case of 
\mbox{$c \, (\mathrm{or} \; d) \to0$} 
where the binary B\"acklund--Darboux transformation 
can reduce 
to an elementary B\"acklund--Darboux transformation. 
Note that 
the Lax matrices in 
(\ref{Lax-Ln2}) and 
(\ref{Lax-Ln3}) are 
both 
\mbox{$(l_1+l_2) \times (l_1+l_2)$} block matrices, i.e., 
they are partitioned in the same 
manner. 

The compatibility condition 
for 
(\ref{gDB3}) and (\ref{gDB4}) 
is given by (a fully discrete 
version 
of) the zero-curvature equation~\cite{AL76,AL77,
Orfa1,Chud2
}
%
\begin{equation}
L_{m+1,n} V_{m,n} = V_{m,n+1} L_{m,n}. 
\label{fd-Lax}
\end{equation}
The Lax matrices $L_{m,n}$ and $V_{m,n}$ 
are of 
the same form 
(cf.~(\ref{Lax-Ln2}) and (\ref{Lax-Ln3})), 
so 
this 
is a matrix re-factorization problem~\cite{Ve03,GoVe04,
Ve07}.   
This 
re-factorization problem 
can be solved explicitly;
that is, for 
the Lax matrices 
originating from a B\"acklund--Darboux transformation 
as given above, 
the two matrices on the left-hand side of (\ref{fd-Lax}) can uniquely
determine the 
two matrices on the right-hand side and vice versa. 
Thus, 
this provides a Yang--Baxter map 
admitting 
the 
Lax (or zero-curvature) 
representation~\cite{GoVe04,SuVe03,Ve07,BoSu08}. 


With the aid of 
a gauge transformation 
\begin{equation}
\Psi_{m,n} = 
\left[
\begin{array}{cc}
 \delta^n d^m I &  \\
 & (-\gamma)^n (-c)^m I \\
\end{array}
\right] 
\Psi'_{m,n}, 
\label{gauge1}
\end{equation}
we can reduce 
the general 
case of \mbox{$\gamma \delta \neq 0$} and 
\mbox{$cd \neq 0$} 
to 
the simpler case of 
\mbox{$\delta =-\gamma=1$} and \mbox{$d =-c=1$} 
after a minor redefinition of the parameters 
and the dependent variables. 
%
The Lax matrices in this 
case 
take 
the same form 
as the standard binary B\"acklund--Darboux transformation~\cite{Sall82}
or the Zakharov--Shabat dressing 
method~\cite{Chud2,ZS79}
up to an 
inessential 
overall factor, 
so they 
admit a 
compact 
expression 
in terms of 
a 
projection 
matrix. 
%

Indeed, 
up to 
an inessential 
$m$-independent 
overall 
factor 
(cf.~(\ref{fd-Lax})), 
we can 
express the Lax matrix 
$L_{m,n}$ 
in the case of 
\mbox{$\delta =-\gamma=1$} 
as 
\begin{align}
L_{m,n} & =  \left[
\begin{array}{cc}
(  \zeta  + \alpha ) I &  \\
  & ( \zeta + \beta) I \\
\end{array}
\right] + (\alpha - \beta )
 \left[
\begin{array}{cc}
 -I & u_{m,n} \\
 v_{m,n} & I \\
\end{array}
\right]^{-1} 
\nonumber \\[1.5mm]
&\propto  
 I + \frac{\alpha - \beta}{\zeta + \beta} P_{m,n}. 
\label{Lax-Ln4}
\end{align}
Here, 
the projection matrix 
$P_{m,n}$ is defined as 
\begin{align}
P_{m,n} &:= 
\left[
\begin{array}{cc}
 I & \\
  & O \\
\end{array}
\right]
+
\left[
\begin{array}{cc}
 -I & u_{m,n} \\
 v_{m,n} & I \\
\end{array}
\right]^{-1} 
\nonumber \\[1mm]
& \hphantom{:}= 
\left[
\begin{array}{cc}
 u_{m,n} v_{m,n} \left( I+u_{m,n} v_{m,n} \right)^{-1}
	& u_{m,n} \left( I+v_{m,n} u_{m,n} \right)^{-1} \\
 v_{m,n} \left( I+u_{m,n} v_{m,n} \right)^{-1} & 
	\left( I+v_{m,n} u_{m,n} \right)^{-1}\\
\end{array}
\right]
\nonumber \\[1mm]
& \hphantom{:}= 
\left[
\begin{array}{c}
 u_{m,n} \\
 I \\
\end{array}
\right] \left( I+v_{m,n} u_{m,n} \right)^{-1}
\left[
\begin{array}{cc}
 v_{m,n} & I \\
\end{array}
\right]
\nonumber \\[1mm]
& \hphantom{:}= 
\left[
\begin{array}{cc}
 -I & u_{m,n} \\
 v_{m,n} & I \\
\end{array}
\right]
\left[
\begin{array}{cc}
 O & \\
  & I \\
\end{array}
\right]
\left[
\begin{array}{cc}
 -I & u_{m,n} \\
 v_{m,n} & I \\
\end{array}
\right]^{-1}, 
\label{Puv}
\end{align}
%
which indeed 
satisfies \mbox{$(P_{m,n})^2 = P_{m,n}$}. 
Note that the correspondence between 
$P_{m,n}$ and 
\mbox{$(u_{m,n}, v_{m,n})$} in (\ref{Puv}) 
is one-to-one.
Owing to the relation
\begin{align}
\left( I + \frac{\alpha - \beta}{\zeta + \beta} P_{m,n} \right) 
\left( I + \frac{\beta - \alpha}{\zeta + \alpha} P_{m,n} \right) 
=I, 
\label{eq3.9}
\end{align}
the inverse of the Lax matrix $L_{m,n}$ 
has 
the same form as $L_{m,n}$
with \mbox{$\alpha \leftrightarrow \beta$}, 
up to 
an inessential 
overall factor. 
Note, incidentally, that 
\mbox{$\left[ \pm (I - 2 P_{m,n}) \right]^2 = I$}. 
Thus, 
using 
the ``spin matrix" 
\mbox{$S_{m,n} := -I + 2 P_{m,n}$}, 
the Lax matrix $L_{m,n}$ can 
be rewritten 
in the form: 
\begin{align}
L_{m,n} 
&\propto  
 I + \frac{\alpha - \beta}{2\zeta + \alpha + \beta} S_{m,n}, 
 \hspace{5mm} (S_{m,n})^2 = I. 
\nonumber 
\end{align}
In the simplest \mbox{$2 \times 2$} case, 
this 
is essentially 
the spatial Lax matrix 
for 
the lattice Heisenberg ferromagnet model~\cite{Ishi82,Skl82} 
(see~\cite{Fad84} and 
\S3.4 of~\cite{2010JPA}
for 
the matrix generalization).

Similarly, we 
can express 
the Lax matrix $V_{m,n}$ in the case of \mbox{$d=-c=1$} 
as 
\begin{align}
V_{m,n} & =  \left[
\begin{array}{cc}
(  \zeta  + a ) I &  \\
  & ( \zeta + b) I \\
\end{array}
\right] + (a-b)
 \left[
\begin{array}{cc}
 -I & q_{m,n} \\
 r_{m,n} & I \\
\end{array}
\right]^{-1} 
\nonumber \\[1.5mm]
&\propto  I 
 + \frac{a - b}{\zeta + b} {\mathscr P}_{m,n}. 
\label{Lax-Ln5}
\end{align}
Here, 
the projection matrix 
${\mathscr P}_{m,n}
$ is defined as 
\begin{align}
{\mathscr P}_{m,n} 
&:= 
\left[
\begin{array}{cc}
 I & \\
  & O \\
\end{array}
\right]
+
\left[
\begin{array}{cc}
 -I & q_{m,n} \\
 r_{m,n} & I \\
\end{array}
\right]^{-1} 
\nonumber \\[1mm]
& \hphantom{:}= 
\left[
\begin{array}{cc}
 -I & q_{m,n} \\
 r_{m,n} & I \\
\end{array}
\right]
\left[
\begin{array}{cc}
 O & \\
  & I \\
\end{array}
\right]
\left[
\begin{array}{cc}
 -I & q_{m,n} \\
 r_{m,n} & I \\
\end{array}
\right]^{-1}, 
\label{Pqr}
\end{align}
which satisfies 
\mbox{$({\mathscr P}_{m,n})^2 = {\mathscr P}_{m,n}$}. 

Substituting (\ref{Lax-Ln4}) and (\ref{Lax-Ln5}) 
into the fully discrete zero-curvature equation (\ref{fd-Lax}), 
we have
\begin{align}
\left( I + \frac{\alpha - \beta}{\zeta + \beta} P_{m+1,n} \right)
\left( I + \frac{a - b}{\zeta + b} {\mathscr P}_{m,n} \right) 
= \left( I + \frac{a - b}{\zeta + b} {\mathscr P}_{m,n+1} \right) 
 \left( I + \frac{\alpha - \beta}{\zeta + \beta} P_{m,n} \right), 
\label{per1}
\end{align}
or equivalently, 
\begin{align}
\left( I + \frac{b - a}{\zeta + a} {\mathscr P}_{m,n} \right) 
 \left( I + \frac{\beta- \alpha}{\zeta + \alpha} P_{m+1,n} \right)
=\left( I + \frac{\beta- \alpha}{\zeta + \alpha} P_{m,n} \right)
\left( I + \frac{b - a}{\zeta + a} {\mathscr P}_{m,n+1} \right), 
\label{per2}
\end{align}
which generalizes 
the re-factorization 
problem 
studied by 
Goncharenko and Veselov~\cite{GoVe04,Ve07}. 
This is an identity in the spectral parameter $\zeta$, 
so 
at 
$O(1/\zeta)$ 
it 
provides the 
conservation law: 
\begin{align}
 (\alpha - \beta) \left( P_{m+1,n} - P_{m,n} \right)
= (a - b) \left( {\mathscr P}_{m,n+1} -{\mathscr P}_{m,n} \right). 
\nonumber
\end{align}
Using 
this 
conservation law, 
we can eliminate 
one of $P_{m+1,n}$, $P_{m,n}$, ${\mathscr P}_{m,n+1}$ and ${\mathscr P}_{m,n}$ 
in the matrix re-factorization problem (\ref{per1}) (or (\ref{per2})). 
Then, it can be solved explicitly
so that 
the map \mbox{$\left( P_{m,n}, {\mathscr P}_{m,n+1} \right) \mapsto 
\left( P_{m+1,n}, {\mathscr P}_{m,n} \right)$} 
and 
the map \mbox{$\left( P_{m,n}, {\mathscr P}_{m,n} \right) \mapsto 
\left( P_{m+1,n}, {\mathscr P}_{m,n+1} \right)$} 
as well as their inverses 
can be expressed in closed form; 
such a map 
represents a similarity transformation 
that 
transforms a projection matrix into a projection matrix 
of the same rank.   
For instance, the 
map 
\mbox{$R_{\{\alpha, \beta\}, \{a,b\}}:\ \left( P_{m,n}, {\mathscr P}_{m,n+1} \right) \mapsto 
\left( P_{m+1,n}, {\mathscr P}_{m,n} \right)$} reads 
\begin{subequations}
\label{P,m+1,n+1}
\begin{align}
&  P_{m+1,n}  = \left( I - \frac{\alpha - \beta}{\alpha - b} P_{m,n}
 - \frac{a - b}{\alpha - b} {\mathscr P}_{m,n+1} \right) P_{m,n}
	 \left( I - \frac{\alpha - \beta}{\alpha - b} P_{m,n} 
	- \frac{a - b}{\alpha - b} {\mathscr P}_{m,n+1} \right)^{-1}, 
\\
& {\mathscr P}_{m,n}  = \left( I - \frac{\alpha - \beta}{\alpha - b} P_{m,n} 
	- \frac{a - b}{\alpha - b} {\mathscr P}_{m,n+1} \right) {\mathscr P}_{m,n+1}
	 \left( I - \frac{\alpha - \beta}{\alpha - b} P_{m,n} 
	- \frac{a - b}{\alpha - b} {\mathscr P}_{m,n+1} \right)^{-1}, 
\end{align}
\end{subequations}
which 
is 
a 
parameter-dependent 
Yang--Baxter map 
acting on the space of 
projection matrices 
and admits the Lax
representation (\ref{per2}); 
the values of the parameters $\alpha, \beta, a$ and $b$ 
are assumed to be 
all different. 
Note that 
the Lax representation is given by 
(\ref{per1}) if we adopt the definition by 
Kouloukas and Papageorgiou~\cite{Kouloukas2009,Kouloukas2011}. 
\begin{center}
\begin{tikzpicture}
\draw (0,0) rectangle (4,4);
\draw (2,-0.1)node[below]{$P_{m,n}, \{\alpha, \beta\}$}; 
\draw (2,4.1)node[above] {$P_{m+1,n}, \{\alpha, \beta\}$} ; 
\draw (-0.1,2)node[left] {${\mathscr P}_{m,n}, \{a,b\}$} ; 
\draw (4.1,2)node[right] {${\mathscr P}_{m,n+1}, \{a,b\}$} ; 
\draw[->,>=stealth] (3.75,0.25)--(0.25,3.75);
\draw (1.85,1.85) node[above right] {$R_{\{\alpha, \beta\}, \{a,b\}}$}; 
\end{tikzpicture}
\end{center}

{\em Remark.} 
The author learned from the recent paper~\cite{Caudrelier22} that 
a slightly reduced form of 
the matrix re-factorization problem (\ref{per1}) (or (\ref{per2})) 
and the resulting 
Yang--Baxter map (\ref{P,m+1,n+1}) 
appeared previously in~\cite{Terng98} (also see 
the relevant paper~\cite{Kouloukas2011}). 

{\em Remark.} 
A map admitting 
the Lax representation
is a Yang--Baxter map 
under 
the 
assumption 
on 
the 
uniqueness of 
factorization of 
the product 
of 
three 
Lax matrices 
(see Kouloukas and Papageorgiou~\cite{Kouloukas2009,Kouloukas2011}). 
In our case, 
this assumption 
means that 
if the relation 
\begin{align}
& \left( I + \frac{\alpha_1 - \beta_1}{\zeta + \beta_1} \widehat{P}_1 \right)
\left( I + \frac{\alpha_2 - \beta_2}{\zeta + \beta_2} \widehat{P}_2 \right)
\left( I + \frac{\alpha_3 - \beta_3}{\zeta + \beta_3} \widehat{P}_3 \right)
\nonumber \\
& = \left( I + \frac{\alpha_1 - \beta_1}{\zeta + \beta_1} P_1 \right)
\left( I + \frac{\alpha_2 - \beta_2}{\zeta + \beta_2} P_2 \right)
\left( I + \frac{\alpha_3 - \beta_3}{\zeta + \beta_3} P_3 \right), 
\label{uniqueness}
\end{align}
holds true, 
where 
\mbox{$( \widehat{P}_j )^2 
= \widehat{P}_j$}, 
\mbox{$(P_j)^2 = P_j$} 
for \mbox{$j=1,2,3$} and \mbox{$\alpha_j, \beta_j \; (j=1,2,3)$} 
are pairwise distinct parameters, 
then we 
have \mbox{$\widehat{P}_j = P_j \; (j=1,2,3)$}. 
This can be checked as follows. 
Using the relation (\ref{eq3.9}), 
we can rewrite (\ref{uniqueness}) as 
\begin{align}
& \left( I + \frac{\beta_1 - \alpha_1}{\zeta + \alpha_1} P_1 \right) 
 \left( I + \frac{\alpha_1 - \beta_1}{\zeta + \beta_1} \widehat{P}_1 \right)
\nonumber \\
& =
\left( I + \frac{\alpha_2 - \beta_2}{\zeta + \beta_2} P_2 \right)
\left( I + \frac{\alpha_3 - \beta_3}{\zeta + \beta_3} P_3 \right)
\left( I + \frac{\beta_3 - \alpha_3}{\zeta + \alpha_3} \widehat{P}_3 \right)
\left( I + \frac{\beta_2 - \alpha_2}{\zeta + \alpha_2} \widehat{P}_2 \right), 
\label{equ3.16}
\end{align}
where the left-hand side can be rewritten as 
\begin{align}
& I + \frac{\beta_1 - \alpha_1}{\zeta + \alpha_1}  \left( P_1 - P_1 \widehat{P}_1 \right)
 + \frac{\alpha_1 - \beta_1}{\zeta + \beta_1} \left( \widehat{P}_1 - P_1 \widehat{P}_1 \right) . 
\nonumber
\end{align}
Thus, by taking the residue at \mbox{$\zeta=-\alpha_1$} and 
\mbox{$\zeta=-\beta_1$}, respectively, in 
(\ref{equ3.16}), 
we have \mbox{$P_1= P_1\widehat{P}_1 = \widehat{P}_1$}. 
In a similar manner, we also have \mbox{$P_2= \widehat{P}_2$} and \mbox{$P_3= \widehat{P}_3$}. 

If (\ref{P,m+1,n+1}) 
is viewed 
as 
a 
system
defined 
on 
the 
two-dimensional lattice,  
the 
parameters 
$\alpha$ 
and $\beta$ 
(resp.~$a$ and $b$)  
can 
be 
arbitrary 
functions 
of the discrete independent 
variable $n$ 
(resp.~$m$). 
Thus, 
(\ref{P,m+1,n+1})
provides 
a new 
fully discrete analog 
of 
(a nonautonomous extension of) 
the principal chiral model~\cite{ZaMi78} 
restricted 
to 
the space of 
projection 
matrices (\mbox{$P^2 = P, \; {\mathscr P}^2 = {\mathscr P}$}): 
\begin{subequations}
\label{P,x,y}
\begin{align}
& 
\frac{\partial P}{\partial \eta}= \frac{f(\eta)}{\alpha(\xi)-a(\eta)
} \left[ P, {\mathscr P} \right], 
\\[2mm]
& \frac{\partial \mathscr P}{\partial \xi} = 
\frac{g(\xi)}{\alpha(\xi)-a(\eta)
} \left[ P, {\mathscr P} \right]. 
\end{align}
\end{subequations}
Here, 
\mbox{$
\left[ P, {\mathscr P} \right] := P {\mathscr P} - {\mathscr P} P
$} 
is 
the commutator. 
When $f(\eta)$ and $g(\xi)$ are purely imaginary and 
$\alpha(\xi)$ and $a(\eta)$ are real, 
the principal chiral model (\ref{P,x,y}) admits 
the 
Hermitian conjugation 
reduction \mbox{$P^\dagger = P$}, \mbox{${\mathscr P}^\dagger = {\mathscr P}$}.
%
%

The Lax-pair representation for 
(\ref{P,x,y}) 
is 
\begin{equation}
\Psi_{\xi} = -\frac{g(\xi)}{\zeta - \alpha(\xi)} P \Psi, \hspace{5mm}
\Psi_{\eta} = -\frac{f(\eta)}{\zeta - a(\eta)} {\mathscr P}  \Psi, 
\label{Lax_PCM}
\end{equation}
where $\zeta$ is a constant spectral parameter. 
Note that 
we can 
normalize $g(\xi)$ and
$f(\eta)$   
to 
nonzero 
constants 
using a 
point transformation of the form 
\mbox{$\xi \to X(\xi), \; \eta \to Y(\eta)$}.
This 
Lax-pair representation 
can 
be 
identified 
with 
(\ref{sG-V}) for two different values of $k$, 
so 
we can 
understand 
the principal chiral model (\ref{P,x,y}) 
as the 
compatibility condition for 
two ``first" negative flows 
of the matrix NLS hierarchy. 

The nonautonomous principal chiral model (\ref{P,x,y})
implies 
the relation 
\begin{align}
& 
\frac{1}{f(\eta)} \frac{\partial P}{\partial \eta} - 
\frac{1}{g(\xi)} \frac{\partial \mathscr P}{\partial \xi} 
= O. 
\label{cons_PCM}
\end{align}
Thus, 
substituting the expression 
(cf.~the Lax-pair representation (\ref{Lax_PCM}))
\[
P = \frac{\alpha(\xi)}{g(\xi)} \Phi_{\xi} \Phi^{-1}, \hspace{5mm}
{\mathscr P} = \frac{a(\eta)}{f(\eta)} \Phi_{\eta} \Phi^{-1}
\]
into (\ref{cons_PCM}), we obtain 
the simpler 
form: 
\[
 \alpha(\xi) \left( \Phi_{\xi} \Phi^{-1} \right)_\eta 
- a(\eta) \left( \Phi_{\eta} \Phi^{-1} \right)_\xi = O. 
\]
It remains an open question 
whether 
this 
model 
is related to 
a more familiar 
nonautonomous chiral model that appears in general relativity 
(see, {\em e.g.},~\cite{DKM-SIGMA} and references therein). 

Let us derive a more explicit 
component form 
of the Yang--Baxter map (\ref{P,m+1,n+1}). 
%
%
In view of the representations 
of the projection matrices 
in (\ref{Puv}) and (\ref{Pqr}), 
we can 
obtain 
from (\ref{P,m+1,n+1}) 
the following relations: 
%
%
\begin{align}
& \left[
\begin{array}{cc}
 -I & u_{m+1,n} \\
 v_{m+1,n} & I \\
\end{array}
\right] 
\left[
\begin{array}{cc}
 X_{m,n} & \\
  & Y_{m,n} \\
\end{array}
\right]
\nonumber \\[2mm] 
=& \left\{ I 
- \frac{\alpha - \beta}{\alpha - b} \left[
\begin{array}{cc}
 u_{m,n} v_{m,n} \left( I+u_{m,n} v_{m,n} \right)^{-1}
	& u_{m,n} \left( I+v_{m,n} u_{m,n} \right)^{-1} \\
 v_{m,n} \left( I+u_{m,n} v_{m,n} \right)^{-1} & 
	\left( I+v_{m,n} u_{m,n} \right)^{-1}\\
\end{array}
\right] - \frac{a - b}{\alpha - b} \times
\right.
\nonumber \\[1mm] & 
\left.  
 \left[
\begin{array}{cc}
 q_{m,n+1} r_{m,n+1} \left( I+q_{m,n+1} r_{m,n+1} \right)^{-1}
	& q_{m,n+1} \left( I+r_{m,n+1} q_{m,n+1} \right)^{-1} \\
 r_{m,n+1} \left( I+q_{m,n+1} r_{m,n+1} \right)^{-1} & 
	\left( I+r_{m,n+1} q_{m,n+1} \right)^{-1}\\
\end{array}
\right] 
\right\}
\left[
\begin{array}{cc}
 -I & u_{m,n} \\
 v_{m,n} & I \\
\end{array}
\right], 
\nonumber
\end{align}
\begin{align}
& \left[
\begin{array}{cc}
 -I & q_{m,n} \\
 r_{m,n} & I \\
\end{array}
\right] 
\left[
\begin{array}{cc}
 Z_{m,n} & \\
  & W_{m,n} \\
\end{array}
\right]
\nonumber \\[2mm] 
=& \left\{ I 
- \frac{\alpha - \beta}{\alpha - b} \left[
\begin{array}{cc}
 u_{m,n} v_{m,n} \left( I+u_{m,n} v_{m,n} \right)^{-1}
	& u_{m,n} \left( I+v_{m,n} u_{m,n} \right)^{-1} \\
 v_{m,n} \left( I+u_{m,n} v_{m,n} \right)^{-1} & 
	\left( I+v_{m,n} u_{m,n} \right)^{-1}\\
\end{array}
\right] - \frac{a - b}{\alpha - b} \times 
\right.
\nonumber \\[1mm] & 
\left.  
 \left[
\begin{array}{cc}
 q_{m,n+1} r_{m,n+1} \left( I+q_{m,n+1} r_{m,n+1} \right)^{-1}
	& q_{m,n+1} \left( I+r_{m,n+1} q_{m,n+1} \right)^{-1} \\
 r_{m,n+1} \left( I+q_{m,n+1} r_{m,n+1} \right)^{-1} & 
	\left( I+r_{m,n+1} q_{m,n+1} \right)^{-1}\\
\end{array}
\right] 
\right\}
\left[
\begin{array}{cc}
 -I & q_{m,n+1} \\
 r_{m,n+1} & I \\
\end{array}
\right],
\nonumber
\end{align}
where $X_{m,n}$, $Y_{m,n}$, $Z_{m,n}$ and $W_{m,n}$ are 
some square matrices. 
By eliminating 
these 
unknown quantities, 
we 
can express 
the Yang--Baxter map $R_{\{\alpha, \beta\}, \{a,b\}}$ 
in component form 
as 
\begin{align}
u_{m+1,n} =&
\left[ u_{m,n} - \frac{a - b}{\beta - b} q_{m,n+1} 
 \left( I+r_{m,n+1} q_{m,n+1} \right)^{-1} \left( I+r_{m,n+1} u_{m,n} \right) \right]
\nonumber \\[0.5mm]
& \mbox{}\times \left[ I - \frac{a - b}{\beta - b}
 \left( I+r_{m,n+1} q_{m,n+1} \right)^{-1} \left( I+r_{m,n+1} u_{m,n} 
	\right) \right]^{-1}
\nonumber \\[2mm]
=&
\left[ \left( \beta - b \right) u_{m,n} \left( I+r_{m,n+1} u_{m,n} \right)^{-1} 
	- \left( a - b \right) q_{m,n+1} 
 \left( I+r_{m,n+1} q_{m,n+1} \right)^{-1} \right]
\nonumber \\[0.5mm]
& \mbox{}\times \left[ \left( \beta - b \right) \left( I+r_{m,n+1} u_{m,n} \right)^{-1} 
 - \left( a - b \right)
 \left( I+r_{m,n+1} q_{m,n+1} \right)^{-1} \right]^{-1}
\nonumber \\[2mm]
=& \left[ \left( \beta - b \right) \left( I+ u_{m,n} r_{m,n+1}  \right)^{-1} 
 - \left( a - b \right)
 \left( I+ q_{m,n+1} r_{m,n+1} \right)^{-1} \right]^{-1}
\nonumber \\[0.5mm]
& \mbox{}\times 
\left[ \left( \beta - b \right)  \left( I+u_{m,n}r_{m,n+1}  \right)^{-1} u_{m,n}
	- \left( a - b \right)  
 \left( I+q_{m,n+1}r_{m,n+1}  \right)^{-1} q_{m,n+1}\right],
%
\nonumber \\[3mm]
v_{m+1,n} =& 
\left[ \left( \alpha - a \right) v_{m,n} \left( I+q_{m,n+1} v_{m,n} \right)^{-1} 
 + \left( a - b \right) r_{m,n+1} 
 \left( I+q_{m,n+1} r_{m,n+1} \right)^{-1}  \right]
\nonumber \\[0.5mm]
& \mbox{}\times \left[ \left( \alpha- a \right) \left( I+q_{m,n+1} v_{m,n} \right)^{-1} 
 + \left( a - b \right)
 \left( I+q_{m,n+1} r_{m,n+1} \right)^{-1} \right]^{-1}, 
\nonumber \\[3mm]
q_{m,n} =& 
\left[ \left( \alpha - a \right) q_{m,n+1} \left( I+v_{m,n} q_{m,n+1} \right)^{-1} 
	- \left( \alpha - \beta \right) u_{m,n} 
 \left( I+v_{m,n} u_{m,n} \right)^{-1} \right]
\nonumber \\[0.5mm]
& \mbox{}\times \left[ \left( \alpha - a \right) \left( I+v_{m,n} q_{m,n+1} \right)^{-1} 
 - \left( \alpha - \beta \right)
 \left( I+v_{m,n} u_{m,n} \right)^{-1} \right]^{-1}
\nonumber \\[2mm]
=&
\left[ \left( \alpha - a \right)  \left( I+q_{m,n+1} v_{m,n} \right)^{-1} 
	- \left( \alpha - \beta \right) 
 \left( I+u_{m,n} v_{m,n} \right)^{-1} \right]^{-1}
\nonumber \\[0.5mm]
& \mbox{}\times \left[ \left( \alpha - a \right) 
	\left( I+q_{m,n+1} v_{m,n} \right)^{-1} q_{m,n+1} 
 - \left( \alpha - \beta \right) 
 \left( I+u_{m,n}v_{m,n}  \right)^{-1} u_{m,n}\right], 
%
%
\nonumber \\[3mm]
r_{m,n} =& 
\left[ \left( \beta - b \right) r_{m,n+1} \left( I+u_{m,n} r_{m,n+1} \right)^{-1} 
 + \left( \alpha - \beta \right) v_{m,n} 
 \left( I+u_{m,n} v_{m,n} \right)^{-1}  \right]
\nonumber \\[0.5mm]
& \mbox{}\times \left[ \left( \beta - b \right) \left( I+u_{m,n} r_{m,n+1} \right)^{-1} 
 + \left( \alpha - \beta \right)
 \left( I+u_{m,n} v_{m,n} \right)^{-1} \right]^{-1}. 
\nonumber
\end{align}
Note that each of the 
expressions for 
$u_{m+1,n}$, $v_{m+1,n}$, $q_{m,n}$ and $r_{m,n}$ 
can be rewritten in 
many different (but nontrivially equivalent) forms, 
as is illustrated for 
$u_{m+1,n}$. 
In fact, 
these expressions can be 
derived 
more 
directly from the definition 
of 
the standard binary B\"acklund--Darboux transformation~\cite{Chud2,Sall82} 
as well as 
its inverse, 
because
the intermediate potentials in 
each projection matrix 
({\em e.g.}, $u_{m,n}$ and $v_{m,n}$ in (\ref{Puv}))
can be expressed explicitly 
in terms of 
the linear eigenfunctions. 
In short, 
a companion 
map 
(see~\cite{ABS04,BoSu08}) 
represents 
the composition law for two binary B\"acklund--Darboux transformations. 

When 
\mbox{$\beta=\alpha^\ast$} and \mbox{$b=a^\ast$}, 
we can impose 
either 
the complex conjugation reduction 
\mbox{$v_{m,n}=\sigma u_{m,n}^\ast$}, \mbox{$r_{m,n+1}=\sigma q_{m,n+1}^\ast$}, 
\mbox{$v_{m+1,n}=\sigma u_{m+1,n}^\ast$}, \mbox{$r_{m,n}=\sigma q_{m,n}^\ast$} 
for square matrices 
or the Hermitian conjugation reduction 
(\mbox{$\ast \to \dagger$}) 
for (generally) rectangular matrices, 
where $\sigma$ is a 
nonzero 
real constant; for simplicity, we do not 
discuss 
a more general 
reduction involving constant Hermitian matrices corresponding to 
the mixed focusing-defocusing case (cf.~\cite{YO2,Ab78,New79,MMP81,Mak82,ZakShul82}). 
In particular, by setting \mbox{$\beta=\alpha^\ast$}, \mbox{$b=a^\ast$}, 
\mbox{$v_{m,n}= u_{m,n}^\dagger$}, \mbox{$r_{m,n+1}= q_{m,n+1}^\dagger$}, 
\mbox{$v_{m+1,n}= u_{m+1,n}^\dagger$}, 
\mbox{$r_{m,n}= q_{m,n}^\dagger$}, 
we obtain the 
parameter-dependent 
Yang--Baxter map 
\mbox{$R_{\alpha, a}:\ \left( u_{m,n}, q_{m,n+1} \right) \mapsto 
\left( u_{m+1,n}, q_{m,n} \right)$}:
\begin{align}
u_{m+1,n} =& \left[ \left( \alpha^\ast - a^\ast \right) 
	\left( I+ u_{m,n} q_{m,n+1}^\dagger  \right)^{-1} 
 - \left( a - a^\ast \right)
 \left( I+ q_{m,n+1} q_{m,n+1}^\dagger \right)^{-1} \right]^{-1}
\nonumber \\[0.5mm]
& \mbox{}\times 
\left[ \left( \alpha^\ast - a^\ast \right) 
	\left( I+u_{m,n} q_{m,n+1}^\dagger \right)^{-1} u_{m,n} 
	- \left( a - a^\ast \right) 
 \left( I+q_{m,n+1} q_{m,n+1}^\dagger \right)^{-1}  q_{m,n+1}\right],
%
\nonumber \\[2mm]
q_{m,n} =& \left[ \left( \alpha - a \right)  
	\left( I+q_{m,n+1} u_{m,n}^\dagger \right)^{-1} 
	- \left( \alpha - \alpha^\ast \right) 
 \left( I+u_{m,n} u_{m,n}^\dagger \right)^{-1} \right]^{-1}
\nonumber \\[0.5mm]
& \mbox{}\times \left[ \left( \alpha - a \right) 
	\left( I+q_{m,n+1} u_{m,n}^\dagger  \right)^{-1} q_{m,n+1}
 - \left( \alpha - \alpha^\ast \right)
 \left( I+ u_{m,n}u_{m,n}^\dagger  \right)^{-1} u_{m,n}\right],
\nonumber 
\end{align}
where the parameters $\alpha$ and $a$ satisfy the conditions 
\mbox{$\alpha \neq a, a^\ast$} and \mbox{$\alpha, a \not \in \mathbb{R}$}.
\begin{center}
\begin{tikzpicture}
\draw (0,0) rectangle (4,4);
\draw (2,-0.1)node[below]{$u_{m,n}, \alpha$}; 
\draw (2,4.1)node[above] {$u_{m+1,n}, \alpha$} ; 
\draw (-0.1,2)node[left] {$q_{m,n}, a$} ; 
\draw (4.1,2)node[right] {$q_{m,n+1}, a$} ; 
\draw[->,>=stealth] (3.75,0.25)--(0.25,3.75);
\draw (2,2) node[above right] {$R_{\alpha, a}$}; 
\end{tikzpicture}
\end{center}

%
Recalling 
that the 
parameter 
$\alpha$ 
(resp.~$a$)  
can 
depend on the discrete independent variable 
$n$ 
(resp.~$m$), 
we obtain a nonautonomous 
system defined on the two-dimensional lattice: 
%
\begin{align}
&  \left( \alpha_n^\ast - a_m^\ast \right) 
	\left( I+ u_{m,n} q_{m,n+1}^\dagger  \right)^{-1} 
	\left( u_{m+1,n} - u_{m,n} \right)
\nonumber \\
&= \left( a_m - a_m^\ast \right)
 \left( I+ q_{m,n+1} q_{m,n+1}^\dagger \right)^{-1} 
	\left( u_{m+1,n} - q_{m,n+1} \right),
%
\nonumber \\[2mm]
& \left( \alpha_n - a_m \right)  
	\left( I+q_{m,n+1} u_{m,n}^\dagger \right)^{-1} 
	\left( q_{m,n+1} - q_{m,n} \right)
\nonumber \\
& = \left( \alpha_n - \alpha_n^\ast \right) 
 \left( I+u_{m,n} u_{m,n}^\dagger \right)^{-1} \left( u_{m,n} - q_{m,n} \right). 
\nonumber 
\end{align}
This 
is 
a fully discrete analog of the principal chiral model for 
Hermitian projection matrices 
in component form: 
\begin{subequations}
\label{rPCM}
\begin{align}
&  \mathrm{i} \frac{\alpha (\xi) - a (\eta)}{\mu (\eta)} 
	 \frac{ \partial u}{\partial \eta}
 = \left( I+ u q^\dagger  \right) 
	\left( I+ q q^\dagger \right)^{-1} \left( u - q \right),
%
\\[1mm]
& \mathrm{i} \frac{\alpha (\xi) - a (\eta)}{\nu (\xi)}
	\frac{ \partial q}{\partial \xi}
 = \left( I+q u^\dagger \right) \left( I+u u^\dagger \right)^{-1} \left( u - q \right). 
\end{align}
\end{subequations}
That is, 
(\ref{rPCM}) 
is obtained from 
the principal chiral model (\ref{P,x,y}) 
by setting 
(cf.~(\ref{Puv0}))
\begin{align}
& f(\eta) = -\mathrm{i} \hspace{1pt} \mu(\eta), \hspace{5mm} 
 g(\xi) = -\mathrm{i} \hspace{1pt} \nu(\xi), 
\nonumber 
\\[2mm]
& P = \left[
\begin{array}{cc}
 I & \\
  & O \\
\end{array}
\right]
+ \left[
\begin{array}{cc}
 -I & u \\
 u^\dagger & I \\
\end{array}
\right]^{-1}, 
\nonumber 
\\[2mm]
& {\mathscr P} = \left[
\begin{array}{cc}
 I & \\
  & O \\
\end{array}
\right]
+ \left[
\begin{array}{cc}
 -I & q \\
 q^\dagger & I \\
\end{array}
\right]^{-1}, 
\nonumber 
\end{align}
%
where \mbox{$\alpha (\xi), \; a(\eta), \; \mu(\eta), \; \nu(\xi) \in \mathbb{R}$}. 
%

%


\section{Concluding remarks}


Auto-B\"acklund transformations 
for a 
continuous integrable system 
provide 
a 
useful 
clue 
to 
obtaining 
its  
proper discretizations, but 
this still remains to be clarified 
for two or more component systems such as the NLS 
system
and their various reductions. 
%
In this paper, 
following the new 
approach 
introduced in our previous paper~\cite{Me2015}, 
we 
obtained  
a new proper space-discretization of the 
matrix NLS 
system, 
%
which 
admits 
the 
Lax-pair representation 
and 
permits 
not only the complex conjugation reduction 
but also 
the Hermitian conjugation reduction between 
the 
two dependent variables 
as in the continuous 
case~\cite{ZS74} (see (\ref{sdNLS5})). 
%
This is in contrast to the 
matrix generalization of the 
Ablowitz--Ladik discretization of the NLS system~\cite{GI82}, 
which 
permits only the complex conjugation reduction 
and 
not the Hermitian conjugation reduction in local form 
(see~\cite{Tsuchi02, DM2010} and references therein). 
Thus, our space-discrete matrix NLS system 
is more 
appropriate 
and physically meaningful; 
indeed, 
it can generate 
proper space-discrete analogs of 
various 
multicomponent NLS 
equations~\cite{Mana74,YO2,MMP81,Mak82,KuSk81,ForKul83,Svi92} 
by considering 
reductions 
analogous to 
the continuous case. 
By changing the time part of the Lax-pair representation, 
we 
can 
also 
obtain 
proper 
discretizations of the matrix 
mKdV system and the matrix sine-Gordon system, 
which 
admit various interesting reductions 
such as the space-discrete vector mKdV equation (\ref{dvmKdV}). 

In our approach, 
we 
reinterpret 
(a slightly generalized version of) 
the binary B\"acklund--Darboux
transformation 
as a discrete spectral problem, 
wherein 
the two intermediate potentials 
appearing 
in 
the Darboux matrix 
are 
used as 
a pair of new 
dependent variables. 
Bianchi's permutability theorem for 
B\"acklund--Darboux transformations 
implies
that 
the number of discrete independent variables 
can be increased consistently
to define 
a multidimensional lattice. 
The consistency condition on 
an elementary 
quadrilateral 
can be solved explicitly, 
providing a 
Yang--Baxter map of the NLS type 
(see the remarks below (\ref{P,m+1,n+1})), 
which 
can be regarded 
as a 
fully discrete 
analog of 
a reduced form of 
the principal chiral model. 
It would be interesting to 
investigate 
whether 
this result 
can be 
related to 
the recent work of 
Caudrelier and 
Q.~C.~Zhang~\cite{Caud2014}. 

%
\section*{Acknowledgments} 
The author 
sincerely 
thanks 
Professor
Baofeng Feng 
for 
his thoughtful 
communication. 
Dr.\ 
Masato Hisakado 
is thanked for 
private communication 
in 1997 
on 
multicomponent generalizations of the 
sine-Gordon equation. 

\appendix
\section{Integrable 
discretization of 
the vector sine-Gordon equation}
\label{append}

In this appendix, we 
present an integrable discretization 
of a vector analog of the sine-Gordon equation~\cite{PR79,EP79}, 
which 
was obtained by the author in the late 1990s but left unpublished. 

We 
start from 
the continuous case and 
consider a 
Lax-pair representation 
of the following form (cf.~(\ref{NLS-U}) and (\ref{sG-V})): 
\begin{subequations}
\label{vsG-Lax}
\begin{align}
& \left[
\begin{array}{c}
 \Psi_1  \\
 \Psi_2 \\
\end{array}
\right]_x 
= \left[
\begin{array}{cc}
-\mathrm{i}\zeta I & -\frac{1}{2\sqrt{c-f}} \mathcal{Q}_x \\
 \frac{1}{2\sqrt{c-f}} \mathcal{R}_x & \mathrm{i}\zeta I \\
\end{array}
\right] 
\left[
\begin{array}{c}
 \Psi_1  \\
 \Psi_2 \\
\end{array}
\right],
\label{vsG-U}
\\[1.5mm]
& \left[
\begin{array}{c}
 \Psi_1  \\
 \Psi_2 \\
\end{array}
\right]_{t_{-1}} 
= \frac{\mathrm{i}}{\zeta}\left[
\begin{array}{cc}
 \sqrt{c-f} \hspace{1pt}I & \mathcal{Q} \\
 \mathcal{R} & -\sqrt{c-f} \hspace{1pt}I \\
\end{array}
\right]
\left[
\begin{array}{c}
 \Psi_1  \\
 \Psi_2 \\
\end{array}
\right]. 
\label{vsG-V}
\end{align}
\end{subequations}
%
%
Here, $c$ is a constant, 
$f$ is a scalar function 
and $\mathcal{Q}$ and $\mathcal{R}$ are square 
matrices. 
The compatibility condition 
for this overdetermined linear system 
provides 
the following 
three 
equations: 
\begin{equation} 
\label{vec_sG1}
\left\{ 
\begin{split}
&  f_x I = \left( \mathcal{Q} \mathcal{R} \right)_x 
	= \left( \mathcal{R} \mathcal{Q} \right)_x, 
\\[0.5mm]
& \left( \frac{1}{\sqrt{c-f}} \mathcal{Q}_x \right)_{t_{-1}} - 4 \mathcal{Q} =O, 
\\[0.5mm]
& \left( \frac{1}{\sqrt{c-f}} \mathcal{R}_x \right)_{t_{-1}} - 4 \mathcal{R} =O.
\end{split}
\right.
\end{equation}
%
To satisfy the 
first equation, 
we 
restrict 
$\mathcal{Q}$ and $\mathcal{R}$ 
to the form: 
\begin{align}
\mathcal{Q} &= q^{(1)} I + \sum_{j=1}^{2M-1} q^{(j+1)} e_j, 
\hspace{5mm}
\mathcal{R} = q^{(1)} I - \sum_{j=1}^{2M-1} q^{(j+1)} e_j, 
\label{reduct-R}
\end{align}
where 
the \mbox{$2^{M-1} \times 2^{M-1}$} matrices
\mbox{$\{ e_1, e_2, 
\ldots, e_{2M-1} \}$} 
are generators of the Clifford algebra; 
they are 
linearly independent 
and 
satisfy 
the anticommutation relations:
\begin{equation}
\{ e_j , e_k \}_+
:=e_j e_k + e_k e_j = -2 \delta_{jk} I. 
\label{a-commute}
\nonumber
\end{equation}
Thus, we have 
\[
\mathcal{Q}\mathcal{R} = \mathcal{R}\mathcal{Q} = \sca{\vt{q}}{\vt{q}} I, 
\]
where \mbox{$\vt{q} = ( q^{(1)}, \ldots, q^{(2M)} )$},  
so the first equation in (\ref{vec_sG1}) is satisfied 
by setting \mbox{$f=\sca{\vt{q}}{\vt{q}}$}. 
Then, 
(\ref{vec_sG1}) 
reduces to the 
vector 
sine-Gordon equation~\cite{PR79,EP79}:\footnote{In constructing 
explicit 
smooth solutions, 
each 
square-root function 
in this appendix 
suffers from 
an intrinsic sign problem. 
Note that (\ref{vsG1}) 
in the scalar case 
reduces to the sine-Gordon equation 
by setting \mbox{$c=1$} and 
\mbox{$\vt{q}=\sin p$}, 
but a sign ambiguity arises
in extracting the square root as 
\mbox{$\sqrt{\cos^2 p} = \pm \cos p$}. 
Such a sign ambiguity 
can be 
resolved by 
expressing 
the dependent variables 
in the Lax pair appropriately 
in terms of trigonometric functions, 
but then the equations of motion become 
complicated in the multicomponent case.} 
\begin{equation}
\left( \frac{1}{\sqrt{c-\sca{\vt{q}}{\vt{q}}}} \vt{q}_x \right)_{t_{-1}} 
	- 4 \vt{q} =\vt{0}, 
\label{vsG1}
\end{equation}
or equivalently, 
\begin{equation}
\vt{q}_{x t_{-1}} 
+ \frac{\sca{\vt{q}}{\vt{q}_{t_{-1}} }}{c-\sca{\vt{q}}{\vt{q}}} \vt{q}_x 
	- 4 \sqrt{c-\sca{\vt{q}}{\vt{q}}} \hspace{1pt} \vt{q} =\vt{0}. 
\nonumber 
\end{equation}
By setting 
\[
-\frac{1}{2 \sqrt{c-\sca{\vt{q}}{\vt{q}}}} \vt{q}_x 
=: \vt{u}, 
\]
which 
changes 
the spectral problem (\ref{vsG-U})
to 
a 
standard 
form, 
we obtain the potential form of 
the vector sine-Gordon equation (\ref{vsG1}): 
\begin{equation}
 \frac{1}{\sqrt{4c - \sca{\vt{u}_{t_{-1}} }{\vt{u}_{t_{-1}}}}} \vt{u}_{t_{-1} x} 
	= 2\vt{u}. 
\label{vsG2}
\end{equation}
More 
information 
on 
multicomponent generalizations 
of the sine-Gordon equation 
can be found in~\cite{PLB96,AncoWolf}
and 
references 
therein. 

Next, we consider 
a discrete-time analog of 
the Lax-pair representation (\ref{vsG-Lax}) 
as given by 
\begin{align}
& \left[
\begin{array}{c}
\Psi_{1,n}  \\
 \Psi_{2,n} \\
\end{array}
\right]_x 
= \left[
\begin{array}{cc}
-\mathrm{i}\zeta I & \varDelta 
	\mathcal{Q}_n -\frac{1}{2\sqrt{c-f_n}} \mathcal{Q}_{n,x} \\
 -\varDelta \mathcal{R}_n 
	+ \frac{1}{2\sqrt{c-f_n}} \mathcal{R}_{n,x} & \mathrm{i}\zeta I \\
\end{array}
\right] 
\left[
\begin{array}{c}
\Psi_{1,n}  \\
 \Psi_{2,n} \\
\end{array}
\right],
\nonumber 
\\[1.5mm]
& 
\left[
\begin{array}{c}
 \Psi_{1, n+1}  \\
 \Psi_{2, n+1} \\
\end{array}
\right]
= \left\{ I + 
\frac{\mathrm{i}\varDelta}{\zeta}\left[
\begin{array}{cc}
 \sqrt{c-f_n} \hspace{1pt}I & \mathcal{Q}_n \\
 \mathcal{R}_n & -\sqrt{c-f_n} \hspace{1pt}I \\
\end{array}
\right] \right\}
\left[
\begin{array}{c}
 \Psi_{1,n}  \\
 \Psi_{2,n} \\
\end{array}
\right], 
\nonumber 
\end{align}
where $\varDelta$ is a 
time-step 
parameter. 
The compatibility condition provides 
\begin{equation} 
\label{vec_dsG1}
\left\{ 
\begin{split}
&  f_{n,x} I = \left( \mathcal{Q}_n \mathcal{R}_n \right)_x 
	= \left( \mathcal{R}_n \mathcal{Q}_n \right)_x, 
\\[0.5mm]
& \frac{1}{\varDelta} \left( \frac{1}{\sqrt{c-f_{n+1}}} \mathcal{Q}_{n+1,x} 
	- \frac{1}{\sqrt{c-f_n}} \mathcal{Q}_{n,x} \right) 
	- 2 \mathcal{Q}_{n+1} - 2 \mathcal{Q}_n =O, 
\\[0.5mm]
& \frac{1}{\varDelta} \left( \frac{1}{\sqrt{c-f_{n+1}}} \mathcal{R}_{n+1,x} 
	- \frac{1}{\sqrt{c-f_n}} \mathcal{R}_{n,x} \right) 
	- 2 \mathcal{R}_{n+1} - 2 \mathcal{R}_n =O.
\end{split}
\right.
\end{equation}
Then, 
by restricting $\mathcal{Q}_n$ and $\mathcal{R}_n$ to the same form as in (\ref{reduct-R}) 
and setting \mbox{$f_n=\sca{\vt{q}_n}{\vt{q}_n}$}, 
(\ref{vec_dsG1})
simplifies to 
a proper 
discrete-time analog of the 
vector 
sine-Gordon equation
(\ref{vsG1}):
\begin{equation}
\frac{1}{\varDelta} \left( 
 \frac{1}{\sqrt{c-\sca{\vt{q}_{n+1}}{\vt{q}_{n+1}}}} \vt{q}_{n+1,x} 
	- \frac{1}{\sqrt{c-\sca{\vt{q}_n}{\vt{q}_n}}} \vt{q}_{n,x} \right) 
 - 2 \vt{q}_{n+1} - 2 \vt{q}_n =\vt{0}. 
\end{equation}
By setting 
\[
\varDelta \vt{q}_n - \frac{1}{2  \sqrt{c-\sca{\vt{q}_n}{\vt{q}_n}}} \vt{q}_{n,x} 
=: 
 \vt{u}_n, 
\]
we obtain a time-discretization of 
the potential 
vector sine-Gordon equation (\ref{vsG2}) as 
\begin{equation}
\frac{1}{\varDelta} \left( 
 \vt{u}_{n+1,x} - \vt{u}_{n,x} \right) 
 =  \sqrt{4c-\frac{1}{\varDelta^2}
	\sca{\vt{u}_{n+1} - \vt{u}_n}{\vt{u}_{n+1} - \vt{u}_n}}
	\left( \vt{u}_{n+1} + \vt{u}_n \right). 
\end{equation}
Essentially the same equation was obtained independently 
by 
Balakhnev and Meshkov~\cite{Bala05,Bala08}
as an auto-B\"acklund transformation
for 
a vector analog of the 
mKdV equation.

\end{document}